\documentclass{jfm}

\usepackage{amsfonts}
\usepackage{amssymb}
\usepackage{amsbsy}
\usepackage{amsmath}
\usepackage{graphicx}
\usepackage{subfigure}
\usepackage{float}
\usepackage{natbib}
\usepackage{array}

\ifCUPmtlplainloaded \else
  \checkfont{eurm10}
  \iffontfound
    \IfFileExists{upmath.sty}
      {\typeout{^^JFound AMS Euler Roman fonts on the system,
                   using the 'upmath' package.^^J}%
       \usepackage{upmath}}
      {\typeout{^^JFound AMS Euler Roman fonts on the system, but you
                   dont seem to have the}%
       \typeout{'upmath' package installed. JFM.cls can take advantage
                 of these fonts,^^Jif you use 'upmath' package.^^J}%
       \providecommand\upi{\pi}%
      }
  \else
    \providecommand\upi{\pi}%
  \fi
\fi

\ifCUPmtlplainloaded \else
  \checkfont{msam10}
  \iffontfound
    \IfFileExists{amssymb.sty}
      {\typeout{^^JFound AMS Symbol fonts on the system, using the
                'amssymb' package.^^J}%
       \usepackage{amssymb}%
       \let\le=\leqslant  \let\leq=\leqslant
       \let\ge=\geqslant  \let\geq=\geqslant
      }{}
  \fi
\fi

\ifCUPmtlplainloaded \else
  \IfFileExists{amsbsy.sty}
    {\typeout{^^JFound the 'amsbsy' package on the system, using it.^^J}%
     \usepackage{amsbsy}}
    {\providecommand\boldsymbol[1]{\mbox{\boldmath $##1$}}}
\fi

\newcommand\dynpercm{\nobreak\mbox{$\;$dyn\,cm$^{-1}$}}
\newcommand\cmpermin{\nobreak\mbox{$\;$cm\,min$^{-1}$}}

\providecommand\bnabla{\boldsymbol{\nabla}}
\providecommand\bcdot{\boldsymbol{\cdot}}
\newcommand\biS{\boldsymbol{S}}
\newcommand\etb{\boldsymbol{\eta}}

\newcommand\Real{\mbox{Re}} 
\newcommand\Imag{\mbox{Im}} 
\newcommand\Rey{\mbox{\textit{Re}}}  
\newcommand\Pran{\mbox{\textit{Pr}}} 
\newcommand\Pen{\mbox{\textit{Pe}}}  
\newcommand\Ai{\mbox{Ai}}            
\newcommand\Bi{\mbox{Bi}}            
\newcommand\bld{\boldsymbol}

\newcommand\ssC{\mathsf{C}}    
\newcommand\sfsP{\mathsfi{P}}  
\newcommand\slsQ{\mathsfbi{Q}} 

\newcommand\hatp{\skew3\hat{p}}      
\newcommand\hatR{\skew3\hat{R}}      
\newcommand\hatRR{\skew3\hat{\hatR}} 
\newcommand\doubletildesigma{\skew2\tilde{\skew2\tilde{\Sigma}}}

\newsavebox{\astrutbox}
\sbox{\astrutbox}{\rule[-5pt]{0pt}{20pt}}
\newcommand{\astrut}{\usebox{\astrutbox}}

\newcommand\GaPQ{\ensuremath{G_a(P,Q)}}
\newcommand\GsPQ{\ensuremath{G_s(P,Q)}}
\newcommand\p{\ensuremath{\partial}}
\newcommand\tti{\ensuremath{\rightarrow\infty}}
\newcommand\kgd{\ensuremath{k\gamma d}}
\newcommand\shalf{\ensuremath{{\scriptstyle\frac{1}{2}}}}
\newcommand\sh{\ensuremath{^{\shalf}}}
\newcommand\smh{\ensuremath{^{-\shalf}}}
\newcommand\squart{\ensuremath{{\textstyle\frac{1}{4}}}}
\newcommand\thalf{\ensuremath{{\textstyle\frac{1}{2}}}}
\newcommand\Gat{\ensuremath{\widetilde{G_a}}}
\newcommand\ttz{\ensuremath{\rightarrow 0}}
\newcommand\ndq{\ensuremath{\frac{\mbox{$\partial$}}{\mbox{$\partial$} n_q}}}
\newcommand\sumjm{\ensuremath{\sum_{j=1}^{M}}}
\newcommand\pvi{\ensuremath{\int_0^{\infty}%
  \mskip \ifCUPmtlplainloaded -30mu\else -33mu\fi -\quad}}

\newcommand\etal{\mbox{\textit{et al.}}}
\newcommand\etc{etc.\ }
\newcommand\eg{e.g.\ }

\title[Stress relaxation in a dilute bacterial suspension]{Stress relaxation in a dilute bacterial suspension: The active-passive transition}

\author[Sankalp Nambiar, Phanikanth S, Nott P.R. and Ganesh Subramanian]%
{Sankalp Nambiar$^1$, Phanikanth S$^2$, \ns Nott P.R.$^2$\break
and Ganesh Subramanian$^1$  \thanks{Email address for correspondence: sganesh@jncasr.ac.in}}

\affiliation{$^1$Engineering Mechanics Unit,
JNCASR, Jakkur, Bangalore 560064, India\\[\affilskip]
$^2$Department of Chemical Engineering, IISc, Bangalore 560012, India}

\date{?; revised ?; accepted ?. - To be entered by editorial office}
\begin{document}

\maketitle

\begin{abstract}
This paper follows a recent article of \cite{nambiar17} on the linear rheological response of a dilute bacterial suspension (e.g. $E.$ $coli$) to impulsive starting and stopping of simple shear flow. Here, we analyse the time dependent non-linear rheology for a pair of impulsively started linear flows - simple shear (a canonical weak flow) and uniaxial extension (a canonical strong flow). The rheology is governed by the bacterium orientation distribution which satisfies a kinetic equation that includes rotation by the imposed flow, and relaxation to isotropy via rotary diffusion and tumbling. The relevant dimensionless parameters are the Peclet number $\Pen \equiv\dot{\gamma}\tau$, which dictates the importance of flow-induced orientation anisotropy, and $\tau D_r$, which quantifies the relative importance of the two intrinsic orientation decorrelation mechanisms (tumbling and rotary diffusion). Here, $\tau$ is the mean run duration of a bacterium that exhibits a run-and-tumble dynamics, $D_r$ is the intrinsic rotary diffusivity of the bacterium and $\dot{\gamma}$ is the characteristic magnitude of the imposed velocity gradient. The solution of the kinetic equation is obtained numerically using a spectral Galerkin method, that yields the rheological properties (the shear viscosity, the first and second normal stress differences for simple shear, and the extensional viscosity for uniaxial extension) over the entire range of $\Pen$. For simple shear, we find that the stress relaxation predicted by our analysis at small $\Pen$ is in good agreement with the experimental observations of \cite{Gachelin2015}. However, the analysis at large $\Pen$ yields relaxations that are qualitatively different. The rheological response in the experiments corresponds to a transition from a nearly isotropic suspension of active swimmers at small $\Pen$, to an apparently (nearly) isotropic suspension of passive rods at large $\Pen$. In contrast, the computations yield the expected transition to a nearly flow-aligned suspension of passive rigid rods at high $\Pen$. We probe this active-passive transition systematically, complementing the numerical solution with analytical solutions obtained from perturbation expansions about appropriate base states. Our study suggests courses for future experimental and analytical studies that will help understand relaxation phenomena in active suspensions.
\end{abstract}


\section{Introduction}
Self propelled micro-swimmers have remained a subject of intense investigations in recent times [\cite{ramaswamy2010, koch2011collective, subprabhu2011, marchetti2013}]. In particular, suspensions of rear-actuated micro-organisms, termed pushers (for instance, the bacteria \emph{E. coli} and \emph{B. subtilis}), exhibit non-trivial behaviour in comparison to their passive counterparts, due to the extensile nature of the intrinsic force dipoles; a key novel feature is the transition to a state of collective motion beyond a critical threshold  [\cite{G2008, saintillan2008, Subkoch2009, Deepak2015, stenhammar2017}]. In the rheological context, observations [\cite{yodh2007, Sokolov2009, gachelin2013, ranga2015, clement2016}] and computations based on a continuum descriptions [\cite{Saintillan2010, saintillan2010extensional, nambiar17, Bechtel2017, saintillan2018}], suggest that pushers tend to decrease the suspension viscosity below that of the solvent alone, and that a sheared suspension of pushers may, in fact, exhibit a negative shear viscosity. However, the focus of the efforts above has been on the steady state rheology.

Recently, experiments have also probed the nature of the stress relaxation in a dilute suspension of \emph{E. coli} subject to an impulsively started shear flow [\cite{Gachelin2015}]. The said experiments were carried out in a sensitive Couette rheometer specifically designed to probe low-viscosity fluids, and involved shearing suspensions of two strains of wild type \emph{E. coli} (RP437 and ATCC9367) for a period of 30 seconds, starting from a state of rest. \cite{Gachelin2015} observed an initial viscous jump followed by an elastic relaxation to the steady-state when the flow was initiated, and a similar viscous fall followed by an elastic relaxation to zero stress when the flow was ceased. The shear rate $\dot{\gamma}$ was varied from as low as 0.02 $s^{-1}$ to 63.4 $s^{-1}$. In terms of the Peclet number $\Pen = \dot{\gamma}\tau$ (which describes the relative importance of the flow induced anisotropy to the randomizing effect of the intrinsic reorientation dynamics, with $\tau$ being the mean run duration of the swimmer), the measured response included both the linear ($\Pen\ll 1$) and the strongly non-linear ($\Pen\gg 1$) regimes. In the linear regime, the elastic relaxation following the initial viscous jump involved a fall in the viscosity; in contrast, for $\Pen\gg 1$, following the viscous jump, the viscosity rose further to the eventual steady state plateau. The transient relaxation of the viscosity in the linear regime was explained by \cite{nambiar17} as being crucially related to the orientational anisotropy of the extensile bacterium force dipoles.

In the present work, our goal is to treat the time dependent response of a dilute sheared suspension of micro-swimmers over the entire range of $\Pen$ for two classes of flows, viz., simple shear and uni-axial extensional flows, with an emphasis on the non-linear regime. While the examination of simple shear is motivated again by the \cite{Gachelin2015} experiments, the consideration of the strong-flow response is motivated by another recent experiment, wherein a novel acoustically-driven microfluidic capillary-breakup extensional rheometer was employed to characterize the extensional viscosity of a suspension of \emph{E. coli} [\cite{ranga2015}]. For simple shear, we determine the time dependent shear viscosity, first and second normal stress differences, whereas for axisymmetric extension, we determine the time dependent extensional viscosity. The paper is organized as follows. In \textsection \ref{sec:stress}, we solve the kinetic equation governing the orientation probability density function for a suspension of slender swimmers, using a Galerkin method after expanding the probability density as a series in surface spherical harmonics. We then evaluate the bulk averaged stress and write down the expression for the relevant rheological quantities. In \textsection \ref{sec:rheo}, we provide numerical results for the transient rheological response in both uni-axial extension and simple shear, the latter over a range of $\Pen$ that roughly correspond to the experiments of \cite{Gachelin2015}. In \textsection \ref{sec:highPe}, an analytical formulation for the swimmer orientation distribution is presented for $\Pen\gg 1$. In this regime, the numerical predictions exhibit a stark disagreement with the results of \cite{Gachelin2015}. Finally, in \textsection\ref{sec:conclusion}, we comment on the implications of the above disagreement, while also examining the various contributions to the active suspension stress, including the recently proposed swim stress contribution [\cite{brady17}]. Appendix \ref{appendix_higherOrder} contains an expansion of the orientation probability density to O($\Pen^3$) for simple shear, and to O($\Pen^2$) for uni-axial extension; the analytical results serve as a useful validation for the numerics. Appendix \ref{appendix_validation} shows the steady state rheological properties, validated against those perviously reported, for both flows considered, and the scalings for the steady-state normal stress differences, in case of simple shear, both, as a function of $\Pen$.

\section{Dilute suspension stress}\label{sec:stress}
In this section, we formulate expressions for the time dependent orientation probability density and the bulk hydrodynamic stress in a swimmer suspension.

\subsection{Orientation probability density}\label{subsec:omega}
The equation governing the probability density $\Omega(\boldsymbol{p}, t)$ for the swimmer orientation, is given by [\cite{Subkoch2009, nambiar17}]:
\begin{equation}
\frac{\partial \Omega}{\partial t}  + \frac{1}{\tau} \Big (\Omega - \int\!\! K({\boldsymbol{p}}|\boldsymbol{p^\prime}) \Omega (\boldsymbol{p^\prime}, t) \mathrm d\boldsymbol{p^\prime} \Big ) - D_r \nabla^2_{\boldsymbol{p}} \Omega = - \bnabla_{\boldsymbol{p}} \!\cdot\! (\dot{\boldsymbol{p}} \Omega).
\label{eq:prob}
\end{equation}
The terms within the brackets in (\ref{eq:prob}) capture the evolution of the orientation distribution due to run-and-tumble dynamics; $K(\boldsymbol{p}|\boldsymbol{p}^\prime) = \beta/(4\pi\sinh\beta)\exp(\boldsymbol{p}\cdot\boldsymbol{p}^\prime)$ denotes the correlation between the pre- and post-tumble orientations, with $\beta$ being the correlation parameter that is zero for random tumbling, and is about 1 for typical wild-type \emph{E. coli} motion of the micro-scale [\cite{Berg93, Subkoch2009}]. The orientational Laplacian arises from the (athermal) rotary diffusion of the bacterium with diffusivity $D_r$. The effect of rotation due to an ambient linear flow is captured by the term on the right side of equation (\ref{eq:prob}), where $\dot{\boldsymbol{p}}$ for slender bodies in an ambient linear flow is given by [\cite{jeffery1922}]
\begin{equation}
\dot{\boldsymbol{p}} = \boldsymbol{\omega} \cdot \boldsymbol{p} + \boldsymbol{E} \cdot \boldsymbol{p} - (\boldsymbol{E}:\boldsymbol{p} \boldsymbol{p})\boldsymbol{p}.
\label{eq:jeffrey}
\end{equation}
Here, $\boldsymbol{\omega}$ and $\boldsymbol{E}$ are the vorticity and rate of strain tensors of the ambient flow field, respectively. In (\ref{eq:prob}) the spatial dependence of $\Omega$ has been neglected, since we consider homogeneous linear flows. A more detailed description of the various terms and their significance is given in \cite{nambiar17}.

We cast (\ref{eq:prob}) in dimensionless form by scaling time with $\tau$, and $\boldsymbol{E}$, $\boldsymbol{\omega}$ with $\dot{\gamma_{0}}$ (the characteristic strength of the imposed velocity gradient), but retain the original notation for brevity, to get
\begin{equation}
\frac{\partial \Omega}{\partial t}  + \Big (\Omega - \int\!\! K({\boldsymbol{p}}|\boldsymbol{p^\prime}) \Omega (\boldsymbol{p^\prime}, t) \mathrm d\boldsymbol{p^\prime} \Big ) - (D_r\tau) \nabla^2_{\boldsymbol{p}} \Omega = - \Pen \bnabla_{\boldsymbol{p}} \!\cdot\! (\dot{\boldsymbol{p}} \Omega).
\label{eq:probND}
\end{equation}

\cite{nambiar17} solved (\ref{eq:probND}) for $\Pen\ll1$, as a regular perturbation expansion in $\Pen$ keeping only the leading term to characterize the linear response. In appendix \ref{appendix_higherOrder}, we extend their solution to higher orders in $\Pen$, and obtain the probability density to $O(\Pen^{2})$ for axisymmetric extensional flow, and to $O(\Pen^{3})$ for simple shear. For arbitrary $\Pen$, we resort to a numerical solution using the Galerkin method [\cite{scheraga1951, scheraga1955, doi1978, strand1987, chen1996rheology}]. The method is based on expressing $\Omega$ as a series in spherical harmonics
\begin{equation}
\Omega(\boldsymbol{p}, t) = \sum_{n=0}^\infty\sum_{m=-n}^n a_{nm}(t) Y_n^m(\boldsymbol{p})
\label{eq:OPDExp}
\end{equation}
where, $Y_n^m$ are the surface spherical harmonics defined by 
\begin{equation}
 Y_n^m(\boldsymbol{p}) \equiv Y_n^m(\theta, \phi) = \sqrt{(2n+1)/(4\pi)(n-m)!/(n+m)!} P_n^m(\cos\theta) \exp(\mathrm{i} m\phi)
\label{eq:SSHarm}
\end{equation}
 and $P_n^m$ are the associated Legendre Polynomials [\cite{abramowitz1970tables}]. The integral kernel in (\ref{eq:probND}) can then be expressed in terms of the Legendre polynomials as
\begin{equation}
K(\boldsymbol{p}|\boldsymbol{p}^\prime) = \sum_{n=0}^\infty A_n P_n(\boldsymbol{p}\cdot\boldsymbol{p}^\prime) = \sum_{n=0}^\infty\sum_{m,=-n}^n \frac{4\pi A_n}{(2n+1)}Y_n^m(\boldsymbol{p})Y_n^{m\ast}(\boldsymbol{p}^\prime),
\label{eq:kernel}
\end{equation}
where we have used the addition theorem of spherical harmonics. 

For axisymmetric extension, $m = 0$, and the expansion reduces to one in Legendre polynomials, $\Omega(\boldsymbol{p}, t) = \sum_n a_n(t) P_n(\cos\theta)$, with $\theta$ being the polar angle measured from the extensional axis. For simple shear flow, $\theta$ and $\phi$ correspond to the polar angle measured from the gradient axis towards the flow-vorticity plane and the azimuthal angle measured from the flow axis in the flow-vorticity plane, respectively [\cite{doi1978}]. It will be seen later that this rather unconventional choice of the polar axis for simple shear allows one to express the flow-induced rotation term in (\ref{eq:probND}) in a finite number of terms involving the Wigner 3j symbols. The numerical solution is further complemented by a large-$\Pen$ analysis in \textsection\ref{sec:highPe}. We now proceed to solve (\ref{eq:probND}) for axisymmetric extensional and simple shear flows.

\subsubsection{Axisymmetric extensional flow}\label{subsubsec:ex}
The velocity field for axisymmetric extensional flow is \textbf{u} = $(-x_{1}/2, -x_{2}/2, x_{3})$. Equation (\ref{eq:jeffrey}) then allows one to calculate the rate of rotation of the bacterium based on the imposed extensional flow. Substituting the expansion for orientation distribution, the rate of rotation by the imposed flow, and the expansion for $K(\boldsymbol{p}|\boldsymbol{p}^\prime)$ in (\ref{eq:probND}), we have:

\begin{eqnarray}
\sum_{n=0}^\infty \left[\frac{d a_n}{d t} + a_n - \frac{4\pi A_n}{(2n+1)} a_n +  (D_r\tau)n(n+1) a_n\right] P_n = \frac{3\Pen}{2}\frac{\dot{\gamma}}{\sin\theta} \frac{d}{d \theta}\left(\sin^2\theta \cos\theta \sum_{n=0}^\infty a_n P_n \right).\nonumber\\
\label{eq:OmegaEx}
\end{eqnarray}
Here, $\dot{\gamma}$($\equiv \dot{\gamma}(t)$) is the non-dimensional time dependent rate of extension. Expressing the trigonometric functions in terms of the Legendre polynomials, and using the latter's recursion and orthogonality relations, results in the following system of linear ordinary differential equations for the series coefficients:

\begin{eqnarray}
&&\frac{d a_n}{d t} + a_n - \frac{4\pi A_n}{(2n+1)} a_n +  (D_r\tau)n(n+1) a_n = (2n+1) \Pen \dot{\gamma} \nonumber \\  && \sum_{n^\prime=0}^\infty a_{n^\prime}\left[ \frac{3}{2} \langle n|2|n^\prime\rangle + \frac{3}{4}(n^\prime+1)\langle n|1|n^\prime+1\rangle - \frac{(n^\prime+1)}{4}(\langle n|0|n^\prime\rangle + 2 \langle n|2|n^\prime\rangle) \right],\nonumber\\ 
\label{eq:OmegaEx3}
\end{eqnarray}
 where the integral of the product of three Legendre polynomials is written compactly in terms of the Wigner 3j symbols [\cite{messiah1958, arfken1999mathematical}],
\begin{equation}
 \int_{-1}^{1} P_l(x) P_m(x) P_n(x) dx \equiv \langle l|m|n\rangle = 2 \left(\begin{array}{ccc}l & m & n\\ 0 & 0 & 0\end{array}\right)\left(\begin{array}{ccc}l & m & n\\ 0 & 0 & 0\end{array}\right).
\label{clebsch_Ex}
\end{equation}

\subsubsection{Simple shear flow}\label{subsubsec:ss}
The flow field for simple shear is \textbf{u} = $(x_{2},0,0)$, whence the only nonvanishing elements of $\boldsymbol{E}$ and $\omega$ are $E_{12} = E_{21} = \dot{\gamma}/2$ and $\omega_{12} = - \omega_{21} = \dot{\gamma}/2$, $\dot{\gamma} \equiv \dot{\gamma}(t)$ now being the non-dimensional time dependent shear rate. Following the same approach as for axisymmetric extension, the system of equations for the $a_{lm}$'s is:
\begin{eqnarray}
\sum_{l, m} \Big[\frac{d\, a_{lm}}{d t}  + && \left( a_{lm} - a_{lm} \frac{4\pi A_l}{(2l+1)} \right) - (D_r\tau) l(l+1) a_{lm} \Big] Y_l^m  = \nonumber\\ 
&& \mbox{} -\Pen \dot{\gamma}  \left[ \left(\frac{2}{3}\sqrt{\frac{4\pi}{5}}Y_2^0 + \frac{\sqrt{4\pi}}{3}Y_0^0\right) \,\sum_{l, m} a_{lm}\mathrm{i} L_y|Y_l^m\rangle \right.\nonumber\\
&& \left.\mbox{} + \sqrt{\frac{2\pi}{15}}\left(Y_2^{-1}+Y_2^{1}\right) \,\sum_{l,m}a_{lm} L_z|Y_l^m\rangle\right.\nonumber\\
&& \left.\mbox{} - 3 \sqrt{\frac{2\pi}{15}}\left(Y_2^{-1}-Y_2^{1}\right) \,\sum_{l,m}a_{lm} Y_l^m  \astrut \right].
\label{eq:coeff}
\end{eqnarray}
The definitions of the angular momentum operators in the above equation ($L_y\{\cdot\}$ and $L_z\{\cdot\}$) are given by (see appendix of \cite{doi1978}) [\cite{messiah1958}]:
\begin{eqnarray}
\it{i}L_y|l,m\rangle &=& 
\frac{1}{2}\sqrt{l(l+1)-m(m+1)}|l,m+1\rangle \nonumber\\
&-&\frac{1}{2}\sqrt{l(l+1)-m(m-1)}|l,m-1\rangle \nonumber\\
\mbox{} L_z|l,m\rangle &=& m|l,m\rangle. \nonumber\\
\label{eq:amoper}
\end{eqnarray}
Finally, using the orthogonality of the spherical harmonics, one obtains:
\begin{eqnarray}
&&\frac{d\, a_{lm}}{d t}  + a_{lm} - a_{lm} \frac{4\pi A_l}{(2l+1)} - (D_r\tau) l(l+1) a_{lm}  = -\Pen \dot{\gamma} \nonumber\\ 
&& \mbox{} \sum_{l^\prime, m^\prime} \left[\bigg\{\frac{1}{3}\sqrt{ \frac{4\pi}{5}} \left(a^{\prime}_{l^\prime m^\prime} \langle l,m|Y_2^0|l^\prime, m^\prime+1\rangle-a^{\prime\prime}_{l^\prime m^\prime}\langle l,m|Y_2^0|l^\prime, m^\prime-1\rangle\right)\right. \nonumber\\
&& \left.\mbox{}\;\;\;\;\;\;\;\;\;\;\; + \frac{\sqrt{4\pi}}{6}\left(a^{\prime}_{l^\prime m^\prime}\langle l,m|Y_0^0|l^\prime, m^\prime+1\rangle +a^{\prime\prime}_{l^\prime m^\prime}\langle l,m|Y_0^0|l^\prime, m^\prime-1\rangle \right) \bigg\} \right.\nonumber\\
&& \left.\mbox{} + \sqrt{\frac{2\pi}{15}} \bigg\{a_{l^\prime m^\prime}(\langle l,m|Y_2^{-1}|l^\prime, m^\prime\rangle \left(m^\prime + 3 \right)  +  \langle l,m|Y_2^{1}|l^\prime, m^\prime\rangle \left(m^\prime - 3 \right)) \bigg\} \astrut \right],
\label{eq:coeff2}
\end{eqnarray}
where, $a^{\prime}_{l^\prime m^\prime} = a_{l^\prime m^\prime} \sqrt{l^\prime(l^\prime+1)-m^\prime(m^\prime+1)}$, $a^{\prime\prime}_{l^\prime m^\prime} = a_{l^\prime m^\prime} \sqrt{l^\prime(l^\prime+1)-m^\prime(m^\prime-1)}$ and $\langle l,m|Y_p^{q}|l^\prime, m^\prime\rangle$ is given by [\cite{doi1978}]:
\begin{equation}
\langle l,m|Y_p^{q}|l^\prime, m^\prime\rangle = (-1)^m \left[\frac{(2l^\prime+1)(2l+1)(2p+1)}{4\pi}\right]^{\frac{1}{2}} \left(\begin{array}{ccc}l & p & l^\prime\\ -m & q & m^\prime\end{array}\right)\left(\begin{array}{ccc}l & p & l^\prime\\ 0 & 0 & 0\end{array}\right).
\label{clebsch_SS}
\end{equation}

We solve the systems (\ref{eq:OmegaEx3}) and (\ref{eq:coeff2}) numerically using the biorthogonal expansion technique [\cite{strand1987}]. For simple shear flow at $\Pen = 100$, $l_{Max} = 64$ corresponding to a total of 4225 terms in the spherical harmonics series. For axisymmetric extension, $l_{Max} = 264$, which was also the actual number of terms.

\subsection{Bulk hydrodynamic stress}\label{subsec:stress}
The deviatoric part of the bacterial contribution to the bulk hydrodynamic stress can be obtained as an orientation space average of the stresslet $\boldsymbol{S}(\boldsymbol{p})$ as:
\begin{equation}
\langle\boldsymbol{\sigma}^b\rangle = n\int \mathrm{d}\boldsymbol{p}\,\, \Omega(\boldsymbol{p}, t) \boldsymbol{S}(\boldsymbol{p}),
\label{bactStress}
\end{equation}
where, $n$ is the bacterial number density. Using the definition of the stresslet in terms of the swimmer force density, we obtain
\begin{equation}
\langle\boldsymbol{\sigma}^b\rangle = n\int \mathrm{d}\boldsymbol{p}\,\, \Omega(\boldsymbol{p}, t) \int_{-L/2}^{L/2} s \left[\boldsymbol{pf}(s) + \boldsymbol{f}(s)\boldsymbol{p} - \frac{2}{3}(\boldsymbol{f}(s)\cdot\boldsymbol{p})\boldsymbol{I}\right]\mathrm{d}s,
\label{bactStress2}
\end{equation}
where the inner integral is over the coordinate $s$ along the axis of the slender swimmer having total length (head + tail) $L$, and $\boldsymbol{f}(s)$ is the linear force density. The force density has two contributions, one arising due to the inextensibility of the swimming bacterium (passive) and another due to its intrinsic activity, that is, $\boldsymbol{f}(s)$ =  $\boldsymbol{f}^{p}(s)$ + $\boldsymbol{f}^{a}(s)$. Substituting the expressions for the two contributions, as given in \cite{nambiar17}, the deviatoric part of the bacterial stress scaled by $\mu \dot{\gamma}$ reduces to:
\begin{eqnarray}
\langle\boldsymbol{\sigma}^b\rangle  = && \frac{\pi (nL^{3})}{6 \ln \kappa}\dot{\gamma}\int \mathrm{d}\boldsymbol{p}\, \Omega(\boldsymbol{p},t)(\boldsymbol{E}:\boldsymbol{pp})\left(\boldsymbol{pp}-\frac{\boldsymbol{I}}{3}\right) \nonumber\\  && - \frac{\alpha}{2M}\frac{nUL^2\tau}{\Pen}\int \mathrm{d}\boldsymbol{p}\, \Omega(\boldsymbol{p},t) \left(\boldsymbol{pp}-\frac{\boldsymbol{I}}{3}\right),
\label{eq:bactstress3}
\end{eqnarray}
where the first term on the right side of the above equation is the passive hydrodynamic contribution, and the second term is the active hydrodynamic contribution, due to the intrinsic force dipoles, with $M$ being the head mobility coefficient and $\alpha$ being the head to total bacterium length.

\subsubsection{Axisymmetric extensional flow}\label{subsubsec:ex2}
On account of axisymmetry, determination of one of the diagonal stress components, $\langle{\sigma}_{E}^b\rangle_{33}$ from (\ref{eq:bactstress3}), is sufficient, and this takes the form:
\begin{eqnarray}
\langle{\sigma}_{E}^b\rangle_{33}  = && \frac{\pi (nL^{3})}{9 \ln \kappa}\dot{\gamma}(t)\int \mathrm{d}\boldsymbol{p}\, \Omega(\boldsymbol{p},t)(P_2(\cos\theta))^2 \nonumber\\  && - \frac{\alpha}{3M}\frac{nUL^2\tau}{\Pen}\int \mathrm{d}\boldsymbol{p}\, \Omega(\boldsymbol{p},t) P_2(\cos\theta).
\label{eq:bactstressEX}
\end{eqnarray}
Using the orthogonality of the Legendre polynomials, the orientation integrals in (2.15) may be written in terms of the coefficients $a_n$, the result being:
\begin{equation}
\langle{\sigma}_{E}^b\rangle_{33} = \frac{4\pi^2 (nL^{3})}{45 \ln \kappa}\dot{\gamma}(t)\left[a_0(t) + \frac{2}{7} \left( a_2(t) + a_4(t) \right)\right]  - \frac{4\pi\alpha (n U L^2\tau)}{15M\Pen} a_2(t),
\label{eq:bactstressEX2}
\end{equation}
with $a_n(t)$'s being the coefficients defined in \textsection\ref{subsubsec:ex}.
\subsubsection{Simple shear flow}\label{subsubsec:ss2}
Here, the lone off-diagonal term, viz. $\langle{\sigma}_{S}^b\rangle_{12}$, gives the shear stress, while the diagonal terms yield the normal stress differences. Proceeding in a manner similar to (\ref{subsubsec:ex2}), the shear viscosity is expressed as
\begin{eqnarray}
\langle{\sigma}_{S}^b\rangle_{12}  = &&\frac{\pi (nL^{3})}{6 \ln \kappa}\dot{\gamma}(t) 
\mbox{} \left[ \frac{2\sqrt{\pi}}{15}a_{0,0}(t) + \frac{2}{21}\sqrt{\frac{\pi}{5}} a_{2, 0}(t) + \frac{1}{7}\sqrt{\frac{2\pi}{15}} \left( a_{2, 2}(t) + a_{2, -2}(t) \right) \right.\nonumber\\
&& \left.\mbox{} -\frac{8\sqrt{\pi}}{105} a_{4, 0}(t) + \frac{2}{21}\sqrt{\frac{2\pi}{5}}\left(a_{4, 2}(t) + a_{4, -2}(t)\right) \right]\nonumber\\
&& \left.\mbox{} + \frac{\alpha (n U L^2\tau)}{2M\Pen}\sqrt{\frac{2\pi}{15}} \bigg [ a_{2, 1}(t) - a_{2, -1}(t) \bigg ]  \astrut \right. .
\label{eq:bactstressSS}
\end{eqnarray}
Similarly, the normal stress differences, $N_1 = (\langle\sigma_S^b\rangle_{11} - \langle\sigma_S^b\rangle_{22})/(\dot{\gamma}\tau)$ ($\equiv (\sigma_{xx} - \sigma_{yy})/(\mu\dot{\gamma}^2\tau)$) and $N_2 = (\langle\sigma_S^b\rangle_{22} - \langle\sigma_S^b\rangle_{33})/(\dot{\gamma}\tau)$ ($\equiv (\sigma_{yy} - \sigma_{zz})/(\mu\dot{\gamma}^2\tau)$) are defined as
\begin{subeqnarray}
	N_1  &=& \frac{\pi (nL^{3})}{6 \ln \kappa}\frac{1}{\Pen}\dot{\gamma}(t) 
	\mbox{} \left[ \frac{1}{3}\sqrt{\frac{\pi}{5}} \left( a_{4,1}(t) - a_{4, -1}(t) \right) - \frac{1}{3}\sqrt{\frac{\pi}{35}} \left( a_{4, 3}(t) - a_{4, -3}(t) \right)\right ]\nonumber\\
	&& \left.\mbox{} + \frac{\alpha (n U L^2\tau)}{2M\Pen^2}\left [2\sqrt{\frac{\pi}{5}} a_{2, 0}(t) -  \sqrt{\frac{2\pi}{15}} \left( a_{2, 2}(t) + a_{2, -2}(t) \right ) \right ]  \astrut \right. ,\\[3pt]
	N_2  &=& \frac{\pi (nL^{3})}{6 \ln \kappa}\frac{1}{\Pen}\dot{\gamma}(t) 
	\mbox{} \left[ \frac{2}{7}\sqrt{\frac{2\pi}{15}} \left( a_{2, -1}(t) - a_{2, 1}(t) \right) + \frac{5}{21}\sqrt{\frac{\pi}{5}} \left( a_{4,-1}(t) - a_{4, 1}(t) \right)\right.\nonumber\\&& \left. \mbox{} + \frac{1}{3}\sqrt{\frac{\pi}{35}} \left( a_{4, -3}(t) - a_{4, 3}(t) \right) \astrut\right ]\nonumber\\
	&& \left.\mbox{} - \frac{\alpha (n U L^2\tau)}{2M\Pen^2}\left [2\sqrt{\frac{\pi}{5}} a_{2, 0}(t) + \sqrt{\frac{2\pi}{15}} \left( a_{2, 2}(t) + a_{2, -2}(t) \right ) \right ]  \astrut \right. .
	\label{eq:bactstressN12}
\end{subeqnarray}

The expressions derived in \textsection\ref{subsubsec:ex2} and \textsection\ref{subsubsec:ss2} are valid for an arbitrary shear-rate history, but in what follows, we restrict ourselves to a study of the response in step shear.

\section{Results and discussion}\label{sec:rheo}
In this section we present results for the case where the flow is initiated at time $t=0$ at a shear rate that subsequently remains constant. Thus, $\dot{\gamma} = H(t)$, the Heaviside function. As discussed in the introduction, the particular time dependence is motivated by the recent study of \cite{Gachelin2015}, wherein bacterial suspensions were subject to an impulsive shear for a finite time interval in a thin-gap Couette device. We only treat the step-up phase, as the step-down phase is expected to be a mirror image of this response [\cite{Gachelin2015, nambiar17}]. The aim is to examine the relaxation of the bulk rheological properties, in both axisymmetric extension (\ref{eq:bactstressEX2}) and simple shear (\ref{eq:bactstressSS} and \ref{eq:bactstressN12}) over the entire range of $\Pen$, including, the particular values examined by \cite{Gachelin2015} for simple shear, thereby enabling a direct comparison. The swimmer parameter set was adopted from \cite{nambiar17}, and is based on the values reported in \cite{Gachelin2015} for the bacteria that were not in a hyper-oxygenated state. We have set $L$ = 8$\mu m$, head length = 2$\mu m$, head volume = 1$\mu m^3$, $U$ = 20$\mu m/s$ and $nL^3$ = $3.43$. The hydrodynamic volume fraction n$L^3$ corresponds to a true volume fraction of 0.67$\%$, considered by \cite{Gachelin2015} (based on the swimmer head volume). The stochastic relaxation parameters, $\tau$ and $D_r$ are set to $1s$ and $0.062 s^{-1}$, respectively [\cite{Berg93, Subkoch2009, nambiar17}]. Most of the results shown below are for the case of random tumbling ($K(\boldsymbol{p}|\boldsymbol{p}^\prime) = 1/(4\pi)$), where $A_0 = 1/(4\pi)$ and $A_n = 0\, \forall\, n\geq1$; we study the effect of correlations (non-zero $\beta$) between the pre and post-tumble orientations on the stress relaxations in \textsection\ref{subsec:tumble}.

The rheology of a passive suspension of axisymmetric Brownian particles under an imposed shear flow has been studied extensively in the classical suspension mechanics literature [\cite{scheraga1951, scheraga1955, leal1972, stewart1972, brenner1974}]. \cite{brenner1974} in his review article, has summarized the steady-state rheology of a dilute suspension of spheroidal particles over a range of the rotary Peclet number  $\Pen_r\equiv \dot{\gamma}D_r$, and over the entire range of aspect ratios; where $D_r$ is the rotary Brownian diffusivity. \cite{strand1987} were the first to examine the time dependent rheology of a suspension of infinitely slender rigid rods in simple shear. We will be comparing our analysis with their work, to compare and contrast the active and passive stress relaxations.

\begin{figure}
\begin{center}
\includegraphics[scale=0.3]{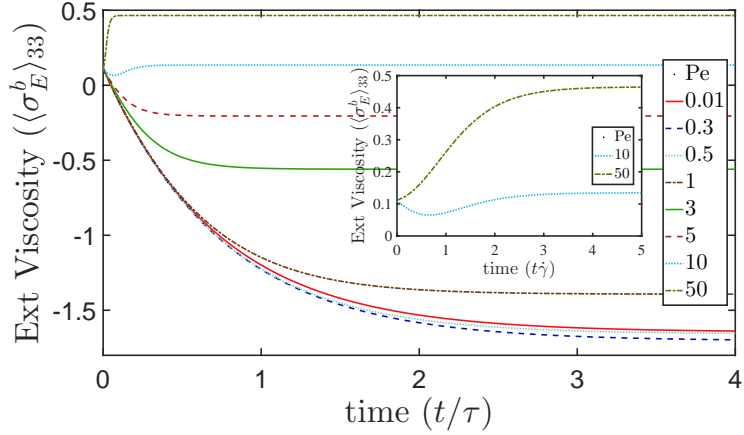}
\end{center}
 \caption{The time dependent bacterial extensional viscosity ($\langle\sigma^b_E\rangle_{33}$), given by (\ref{eq:bactstressEX2}), plotted for $\Pen \sim 0.01-50$. The inset consists of the extensional viscosity for $\Pen = 10$ and $\Pen = 50$, with the time being scaled by the inverse rate of extension.}
 \label{fig:ext_visc}
\end{figure}

\begin{figure}
\begin{center}
\includegraphics[scale=0.3]{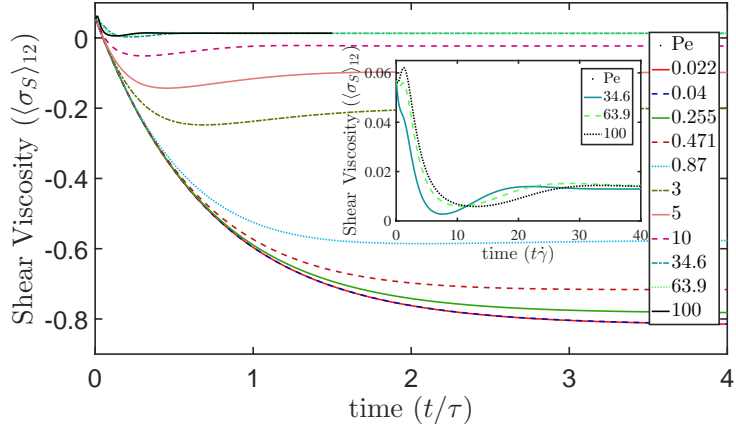}
\end{center}
 \caption{The time dependent shear viscosity ($\langle\sigma_S\rangle_{12}$), given by (\ref{eq:bactstressSS}), plotted for $\Pen \sim 0.022-100$. The inset consists of the shear viscosity, for $\Pen = 34.6$, $\Pen = 63.9$ and $\Pen = 100$, plotted as a function of the time scaled by the inverse shear rate.}
 \label{fig:ss_visc}
\end{figure}

In figure \ref{fig:ext_visc}, we plot the bacterial contribution to the time dependent extensional viscosity for different $\Pen$. For any $\Pen$, the flow distorts the isotropic distribution, creating an excess of orientations in the vicinity of the extensional axis. For  $\Pen \ll 1$, the approach towards the anisotropic steady state occurs in a time of $O(\tau)$, and the oriented extensile swimmer dipoles reinforce the imposed extension, implying a reduction in the total viscosity (below that of the solvent). However, with increasing $\Pen$, the passive contribution to the stress begins to increase and is eventually dominant, implying a steady-state extensional viscosity higher than that of the solvent. Therefore, the steady-state extensional viscosity plateau, starts off being negative at small $\Pen$, passes through a minimum at $\Pen \sim 0.3$, at changes sign at $\Pen\sim 10$. Although the said plateau is positive at $\Pen = 10$, implying a dominant passive stress component, the relaxation to this steady state is nevertheless non-monotonic, implying a non-trivial effect of activity on the transient even at large $\Pen$ (see inset of figure \ref{fig:ext_visc}). For $\Pen = 50$ or larger, however, the active component has a negligible effect at all times, and a monotonic relaxation prevails. Note that as $\Pen$ increases from 0.01 to 50, in figure \ref{fig:ext_visc}, the characteristic relaxation time reduces from being O($\tau$) when $\Pen\ll 1$ to O($\dot{\gamma}^{-1}$) when $\Pen\gg 1$, as per expectation.

Figure \ref{fig:ss_visc} contains only the bacterial contribution to the time dependent viscosity of the swimmer suspension in response to an imposed simple shear (the solvent contribution is unity). The collapse of the relaxation curves for $\Pen$ up to 0.075 is indicative of a linear response regime ($\Pen\ll 1$). An explanation similar to that for extensional flow above can be adapted to simple shear. For $\Pen\ll 1$, the flow  results in an ehnancement in the probability of swimmers oriented around the extensional axis (at 45$^\circ$ to the flow axis); the dipoles associated with the oriented swimmers lead to the reduction in viscosity [\cite{nambiar17}]. As $\Pen$ increases, there is a transition to a non-linear response, with the bacterial stress now being a strong function of the applied shear rate. With increasing $\Pen$, the peak of the swimmer orientation distribution is rotated by the ambient vorticity, leading to a progressive alignment with the flow axis. The misalignment of the intrinsic force dipoles with the extensional axis leads to a shear thickening behavior until $\Pen\approx$ 10 (it is shown in appendix \ref{appendix_higherOrder} that a small-$\Pen$ expansion of the orientation probability density, at O($\Pen^3$), captures the shear rate dependence of the viscosity until $\Pen\approx$ 0.3). Although the steady viscosity is shear thickening for $\Pen\in [3, 10]$, a non-monotonic transient in the viscosity is observed, that is, reaching a mininum and then rising to a plateau higher than that in the linear regime. This is due to the active component, which continues to remains important at moderate $\Pen$, as the bacteria are transiently aligned with the extensional axis on (short) time scales of O($\dot{\gamma}^{-1}$). Further increase in $\Pen$ leads to a transition, from an active stress-dominated regime with a shear-thickening rheology at low to moderate $\Pen$, to a passive stress-dominated regime, which leads to an eventual shear thinning behavior (see figure \ref{high_pe_leading}b). The latter regime is similar to that reported by \cite{strand1987} for a suspension of passive rigid rods. This transition occurs over an interval of $\Pen$ between 10 and 30, and a manifestation of this transition is the emergence of a peak for short times owing to the passive hydrodynamic contribution associated with swimmers momentarily aligned with the extensional axis (see inset of figure \ref{fig:ss_visc}). The characteristic relaxation time reduces from O($\tau$) for $\Pen\ll 1$ to O($\dot{\gamma}^{-1}$) when $\Pen\gg 1$, as in extensional flow. The reader is directed to \textsection\ref{sec:highPe} for further discussion on the $\Pen\gg 1$ regime.

\begin{figure}

\begin{tabular}{p{1.25cm} p{1cm} p{1cm}}
 & \includegraphics[width=0.7\textwidth]{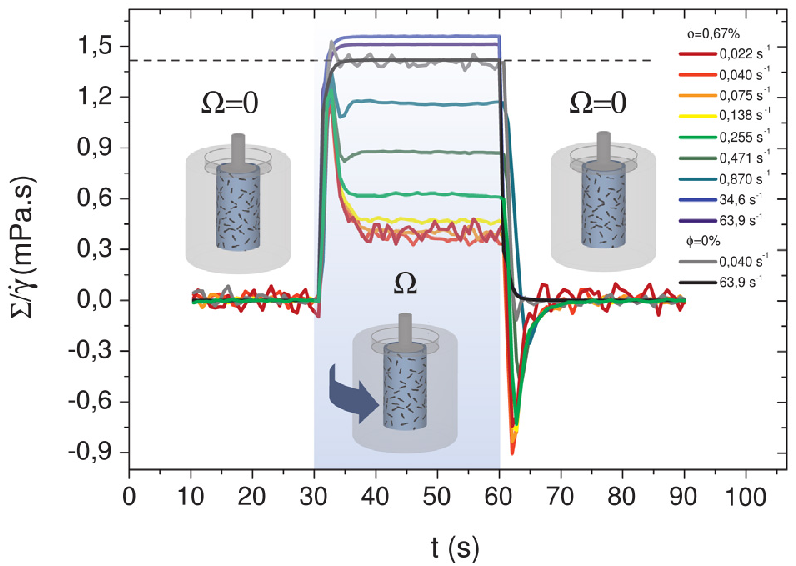} &\\
\end{tabular}
\begin{center}
 (a)
\end{center}

\begin{center}
  \includegraphics[width=0.8\textwidth]{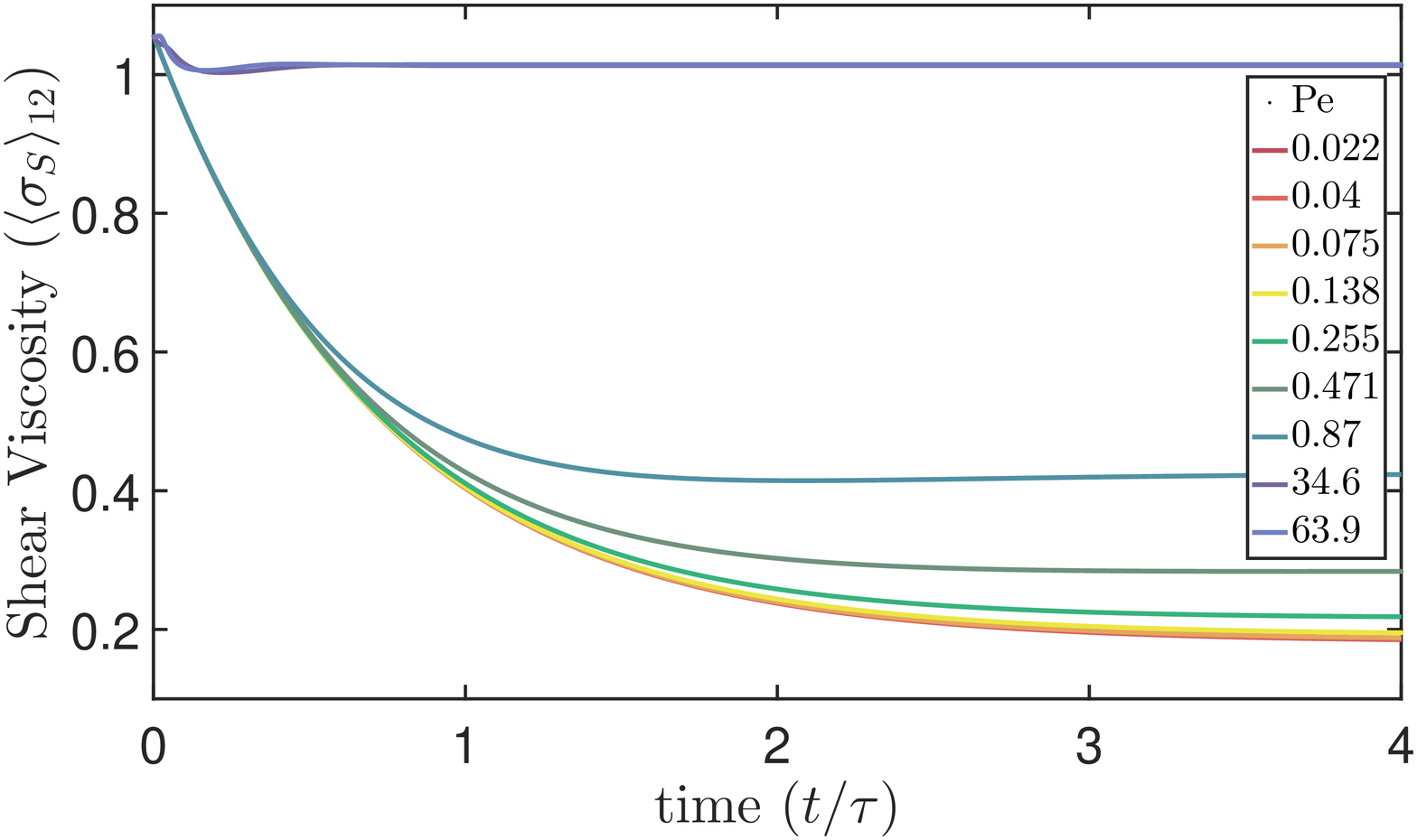}\\
  (b)
\end{center}

 \caption{The time dependent shear viscosity (a) Reported in \cite{Gachelin2015} (Reproduced with permission), and (b) given by (\ref{eq:bactstressSS}), with the solvent contribution included, plotted for $\Pen \sim 0.022-63.9$. For E. coli, the run duration $\tau \approx 1s$, and hence $\Pen$ equals the shear rate $\dot{\gamma}$ \(\text{in } s^{-1}\)). For \emph{E. coli}, the run duration is $\tau\approx 1$s, and hence $\Pen$ equals the shear rate $\dot{\gamma}$ (in s$^{-1}$)}
 \label{exp_theo_comp}
\end{figure}

There are important differences between the experimental findings of \cite{Gachelin2015} and our results in figure \ref{fig:ss_visc} that need to be highlighted. Figure \ref{exp_theo_comp}a and b shows the stress relaxations measured by \cite{Gachelin2015} alongside our theoretical predictions ones for the same values of $\Pen$, with the solvent contribution included in the latter. There is good agreement for the $\Pen$ range  corresponding to the linear regime in the two cases, although the effects of non-linearity appear at smaller $\Pen$ in the experiments. While the observations indicate a pronounced minimum in the time dependent viscosity for $\Pen = 0.471$ and 0.87, this behavior is observed in our computations only starting at $\Pen \approx 3$ (not shown). The more important divergence is the magnitude of the steady state plateaus. The observations indicate a rapid rise in the plateau values for $\Pen > 0.075$ (in the non-linear regime), with the plateau, for $\Pen = 0.87$ being nearly 82$\%$ of the initial viscous jump. In contrast, the computed relaxation at $\Pen = 0.87$, apart from being monotonic, saturates at a value that is only 42$\%$ of the initial viscous jump. More puzzlingly, at the two highest values of $\Pen$ (34.6 and 63.9), the observed relaxation is a monotonic rise to a plateau stress that is above the initial viscous jump. This response mimics a suspension of orientable passive Brownian particles at low $\Pen$; that is, a viscous jump followed by elastic relaxation to a steady-state above the solvent Newtonian plateau. This behaviour is contrary to intuition, as at large $\Pen$ we expect the flow-aligning effect of the imposed shear (and the resulting decrease in the viscosity) to overwhelm any relaxation to isotropy. 

Another point of divergence between our analysis and the observations of \cite{Gachelin2015} is with regard to the relaxation time to the steady state viscosity. The relaxation times remain comparable in the experiment for the entire range of $\Pen$'s examined; it is about 3.2s at low $\Pen$, and slightly above $1s$ at the highest $\Pen$. This observation is difficult to explain. In a dilute bacterial suspension (the assumption of diluteness here being reasonable, given that the true volume fraction in \cite{Gachelin2015} was around 0.67$\%$), the primary relaxation is determined by the faster of the two competing effects, namely the stochastic relaxation mechanisms (run-and-tumble dynamics or rotary diffusion) and flow-induced orientation. This implies that, for their largest shear rate of $\dot{\gamma} = 63.9s^{-1}$, the expected relaxation time is 0.016s. Therefore, the observed relaxation time of $\sim 1$s is higher by a factor of 60! Consideration of a distribution of run times, did account for slightly larger relaxation times observed in the linear response regime [\cite{nambiar17}], but cannot account for the discrepancy in the non-linear response. A possible explanation for the large observed relaxation times is that at the volume fraction considered, the effective volume fraction is $nL^3$ = 3.43, at which the suspension has transitioned to a regime of collective motion [\cite{Deepak2015, nambiar17}].

\begin{figure}
\begin{center}
\subfigure[\hspace{-0.2cm}]{\includegraphics[scale=0.3]{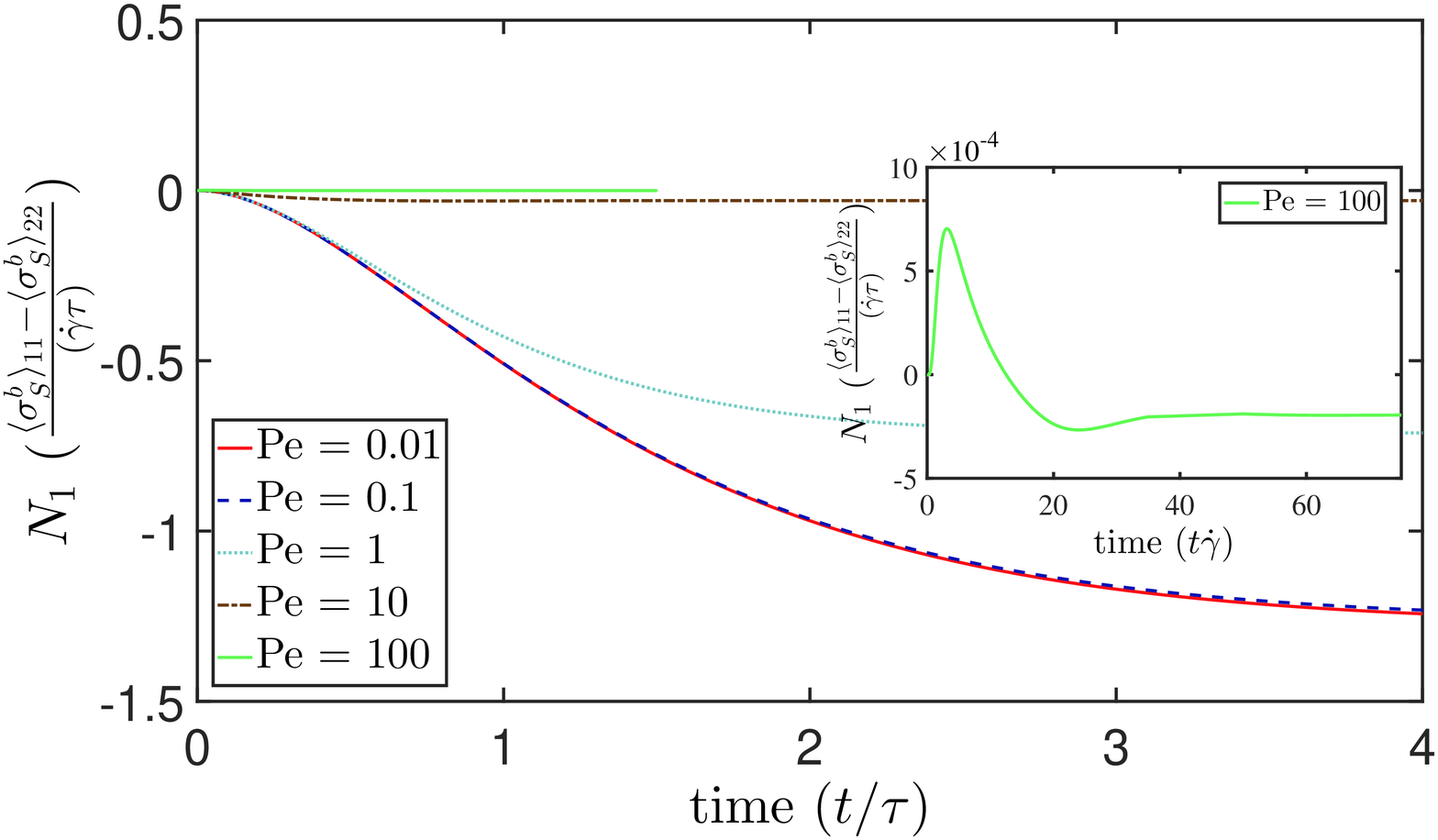}}
\subfigure[\hspace{-0.2cm}]{\includegraphics[scale=0.3]{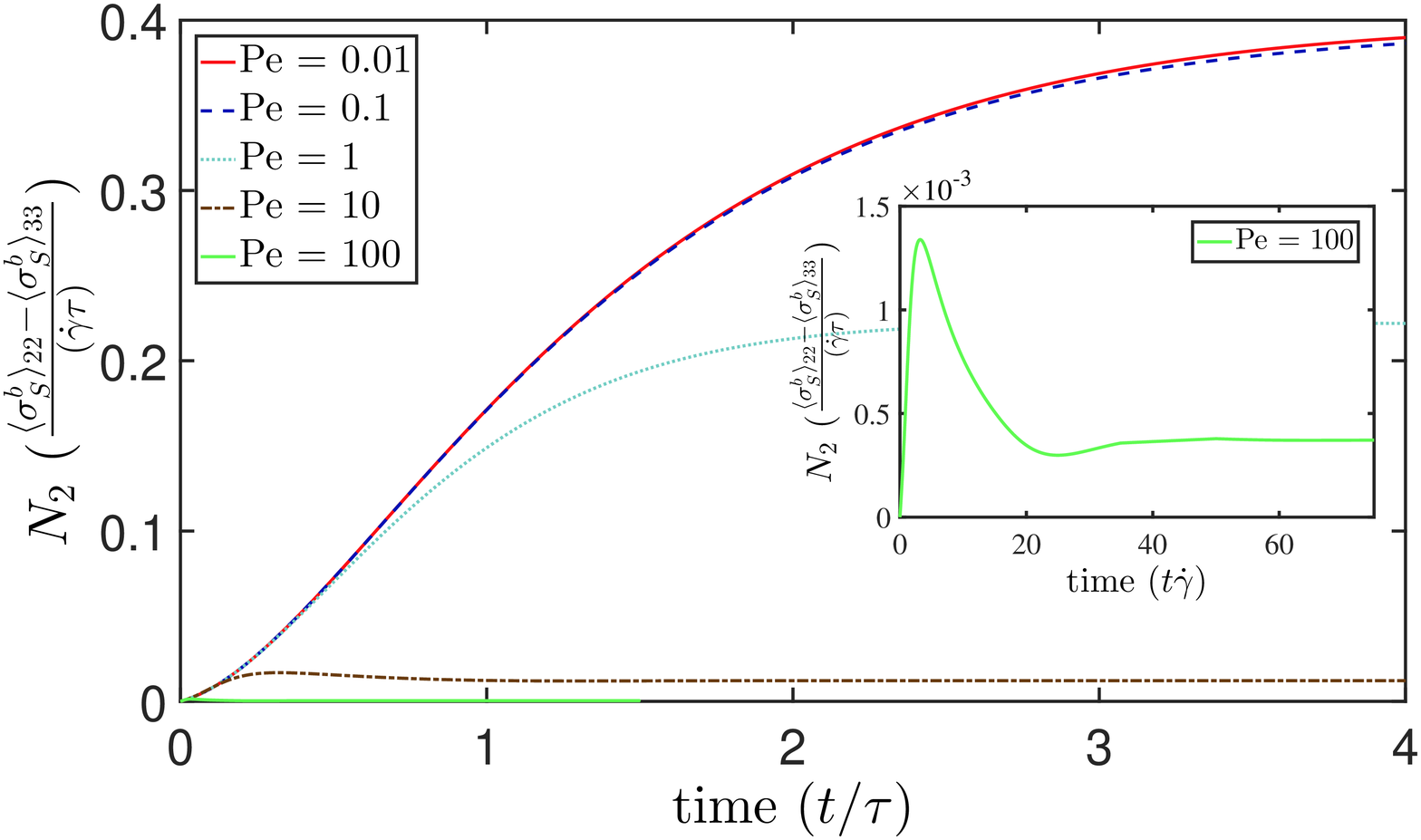}}
\end{center}
 \caption{The time dependent bacterial normal stress differences, plotted for $\Pen \sim [0.01, 100]$. a) $N_{1}$ b) $N_{2}$.}
 \label{exp_num_NSD}
\end{figure}

While the experiments of \cite{Gachelin2015} measured only the shear viscosity, we will also look at the normal stress differences in what follows, since they are representative of the non-trivial elastic signatures associated with pusher suspensions, and might be responsible for triggering elastic instabilities (similar to polymer solutions) in other inhomogeneous flow settings. We first interpret the steady-state results. As shown in appendix \ref{appendix_validation} (figures \ref{saintillan_comparison2}b and c), the first ($N_1$) and the second normal stress ($N_2$) coefficients begin from a low-$\Pen$ plateau, eventually asymptoting to zero with appropriate power-law scalings for $\Pen\rightarrow\infty$. These trends are consistent with the results of \cite{Saintillan2010}.

The temporal variations of the first and second normal stress differences (NSD) for different $\Pen$ are shown in figures \ref{exp_num_NSD}a and \ref{exp_num_NSD}b. When $\Pen\ll 1$, the dominant contributions to both $N_1$ and $N_2$ comes from the active component. Therefore, in this limit, the NSD have signs opposite to those expected for passive suspensions ($N_1$ is negative and $N_2$ is positive), the relaxation to steady state occuring on a time scale of O($\tau$). With increase in $\Pen$, the time scale for the relaxation of the NSD eventually shrinks to O($\dot{\gamma}^{-1}$). Following the same reasoning as for the shear viscosity, one expects the passive component to be dominant when $\Pen\gg1$, implying a sign reversal of the NSD for sufficiently large $\Pen$ [\cite{strand1987, brenner1974}]. For $\Pen = 100$, the peak $N_1$, attained at a time of O($\dot{\gamma}^{-1}$), is indeed positive, despite the eventual plateau being negative. Interestingly, for $N_2$ we observe no sign change for the chosen values of the swimmer parameters. Further discussion on the large-$\Pen$ scaling behaviour of the NSD is given in appendix \ref{appendix_validation}.

\subsection{Effect of correlation in tumbling on the viscosity}\label{subsec:tumble}

In (\ref{eq:probND}), the tumbling kernel $K(\boldsymbol{p}|\boldsymbol{p}^\prime) = \beta\exp(\beta\boldsymbol{p}\cdot\boldsymbol{p}^\prime) /(4\pi\sinh\beta)$ [\cite{Subkoch2009}], is a function of the parameter $\beta$, going from 1/$4\pi$, for $\beta = 0$ (random tumbling) to $\delta(\boldsymbol{p} - \boldsymbol{p}^\prime)$ for $\beta\rightarrow\infty$. The results shown thus far have all been for random tumbling. To study the role of correlation of adjacent tumbles we consider the cases $\beta = 0$ and 1, the latter being consistent with the run-and-tumble statistics of wild-type \emph{E. coli} [\cite{Berg93, Subkoch2009}]. The main observation is that correlated tumbling has no effect on the qualitative nature of the stress relaxation (not shown). Quantitatively, correlated tumbling leads to a marginally slower relaxation of the orientation distribution. This leads to a stronger negative stress at low $\Pen$ (until about 0.5), and a greater positive stress at higher $\Pen$ (greater than about 10). This, together with the differing dominant contributions to the stress, active at low $\Pen$ and passive at large $\Pen$, accounts for the change in the sign of the correlation-induced correction to the random-tumbling plateaus.

\section{The swimmer distribution for $\Pen\gg 1$}\label{sec:highPe}

In this section, we analytically determine  $\Omega(\boldsymbol{p}, t)$, and thence, the bacterial stress $\langle\sigma^b\rangle$, for  $\Pen\gg 1$. We only consider the passive stress component, given that active stress scales as O($1/\Pen$), and does not affect the relaxation for $\Pen\gg 1$. The evolution of $\Omega(\boldsymbol{p}, t)$ in this limit is only due to the imposed flow. Scaling time by $\dot{\gamma}^{-1}$, (\ref{eq:prob}) takes the form:

\begin{equation}
\frac{\partial \Omega}{\partial t}  + \left(\frac{1}{\Pen}\right) \Big (\Omega - \int\!\! K({\boldsymbol{p}}|\boldsymbol{p^\prime}) \Omega (\boldsymbol{p^\prime}, t) \mathrm d\boldsymbol{p^\prime} \Big ) - \frac{1}{\Pen_r} \nabla^2_{\boldsymbol{p}} \Omega = -  \bnabla_{\boldsymbol{p}} \!\cdot\! (\dot{\boldsymbol{p}} \Omega),
\label{eq:probNDHigh}
\end{equation}
where $\Pen_r \equiv \Pen/(\tau D_r) = \dot{\gamma}/D_r$ is now the relevant rotary Peclet number. The run-and-tumble and the rotary diffusion terms are now both O(1/$\Pen$), for a fixed $\tau D_r$, when the angular scale characterizing the orientation probability density is O(1), as is the case for the initial isotropic base state. Thus, at leading order, the orientation distribution is merely convected, on time scales of O($\dot{\gamma}^{-1}$), from an initial isotropic state to being aligned along the extensional axis for axisymmetric extension, and with the flow axis in case of simple shear.

For longer times, stochastic relaxation mechanisms do become relevant. Rotary diffusion becomes important first, due to the large gradients in orientation space that develop in the neighbourhood of the extensional (axisymmetric extension) and flow (simple shear) axes. Scaling arguments suggest that, on a time scale of O($\Pen_r^{1/3}$), there exists a boundary layer, with an angular extent of O($\Pen_r^{-1/3}$), around the said axes, where rotary diffusion and convection are in balance. A detailed analysis of this balance, although possible, is beyond the scope of the present discussion for multiple reasons. The first is, of course, the profound disagreement between the theoretical predictions and the experimental observations [\cite{Gachelin2015}], for $\Pen\gg 1$, that has already been highlighted above. Second, the nature of the large-$\Pen$ relaxation sensitively depends on the presence of rotary diffusion. Although not reported here, we find the large-$\Pen$ stress relaxation for swimmers that exhibit pure run-and-tumble dynamics to be very different. In what follows, we therefore only consider the leading order convective analysis, and use the results to rationalize the high-$\Pen$ numerical results detailed in \textsection\ref{sec:rheo}.

\subsection{Outer convective solution ($\Pen\rightarrow\infty$)}\label{subsub:extHighPeOuter}

Beginning from an isotropic distribution $\Omega(\boldsymbol{p}, 0) = 1/(4\pi)$, the evolution of the swimmer orientation probability density is different for axisymmetric extension and simple shear. In the former case, the orientation distribution is stretched towards alignment with the extensional axis. However, in the latter case, the alignment with the extensional axis (at $45^\circ$) is a transient phenomenon, occuring on time scales of O(1), following which the orientation distribution is rotated towards the flow axis while also being stretched at the same time. Therefore, in addition to the obvious advantage of azimuthal angle independence  (mentioned in \textsection\ref{sec:stress}), the analysis for axisymmetric extension is also simpler since one only needs to consider stretching of $\Omega(\boldsymbol{p}, t)$ with the peak probability always along the extensional axis. At leading order, one needs to solve for the unsteady convective evolution  in (\ref{eq:probNDHigh}), given by:
\begin{equation}
\frac{\partial \Omega_o}{\partial t} + \bnabla_{\boldsymbol{p}} \!\cdot\! (\dot{\boldsymbol{p}} \Omega_o) = 0.
\label{eq:probNDextHigh}
\end{equation}
Here, $\Omega_o$ refers to the outer solution.

\subsubsection{Axisymmetric extensional flow}\label{subsec:extHighPe}

Choosing the coordinate system of \textsection \ref{subsubsec:ex}, substituting for $\dot{\boldsymbol{p}}$, and setting $\cos\theta = x$, simplifies (\ref{eq:probNDextHigh}) to give:
\begin{equation}
\frac{\partial \Omega_o}{\partial t} + \frac{3\hat{\dot{\gamma}}}{2}x(1-x^2)\frac{\partial \Omega_o}{\partial x} + \frac{3\hat{\dot{\gamma}}}{2} (1-3x^2)\Omega_o = 0.
\label{eq:probNDextHigh3}
\end{equation}
The above equation can be solved by using the method of characteristics. Denoting the factor multiplying $\partial\Omega_o/\partial x$ as $\mathrm{d} x/\mathrm{d}t$, one has the following pair of ODE's to solve:
\begin{subeqnarray}
  \frac{\mathrm{d} x}{\mathrm{d} t}  &=& \frac{3\hat{\dot{\gamma}}}{2}x(1-x^2),\\[3pt]
  \frac{\mathrm{d} \Omega_o}{\mathrm{d} t} &=& -\frac{3\hat{\dot{\gamma}}}{2}(1-3x^2)\Omega_o ,
  \label{eq:probNDextHigh4}
\end{subeqnarray}
where $x$ in (\ref{eq:probNDextHigh4}b) is now a function of time. The solution is given by:
\begin{equation}
\Omega_o = \frac{1}{4\pi}\left(\frac{1}{1-\sin^2\theta(1-\exp(3t))}\right)^{\frac{3}{2}}\exp(3t).
\label{eq:probNDextHigh5}
\end{equation}
The factor $\exp(3t)$ implies an exponential increase of $\Omega_o(\boldsymbol{p}, t)$ along the extensional axis. Conservation of probability implies that the width of the distribution decreases as $\exp(-3t/2)$, so that, in the absence of stochastic relaxation, $\Omega_o(\boldsymbol{p}, t)$ approaches a Dirac delta-function for $t\rightarrow\infty$.

On using (\ref{eq:probNDextHigh5}) and only considering the passive component of the stress, relevant for determining the viscosity at large $\Pen$, one obtains
\begin{eqnarray}
  \langle\sigma_E^b\rangle_{33}^p|_{outer} = \frac{\pi (nL^3)}{18\ln\kappa} & & \left[\frac{(\exp(3t)+2)(8\exp(3t)+1)}{4(\exp(3t)-1)^2} \right.\nonumber\\ 
&& \left. \mbox{} - 3\exp(3t)\frac{(5\exp(3t)+4)\tan^{-1}[(\exp(3t)-1)^{1/2}]}{4(\exp(3t)-1)^{5/2}} \astrut \right].
\label{eq:extPassiveStress2}
\end{eqnarray}
In figure \ref{high_pe_leading}a, we plot the extensional viscosity obtained numerically in \textsection\ref{subsec:stress} along with (\ref{eq:extPassiveStress2}), for $\Pen_r = 100$, 200 and 300. The analytical and numerical evolutions are largely in agreement. The numerical plateaus lie slightly below the analytical one owing to the effects of rotary diffusion, which eventually arrests the convective evolution, yielding an $\Omega$ that is a narrow Gaussian with an O($\Pen_r^{-1/3}$) angular spread about the extensional axis, instead of the delta function. 

\subsubsection{Simple shear flow}\label{subsec:ssHighPe}

\begin{figure}
\begin{center}
\subfigure[\hspace{-0.2cm}]{\includegraphics[scale=0.3]{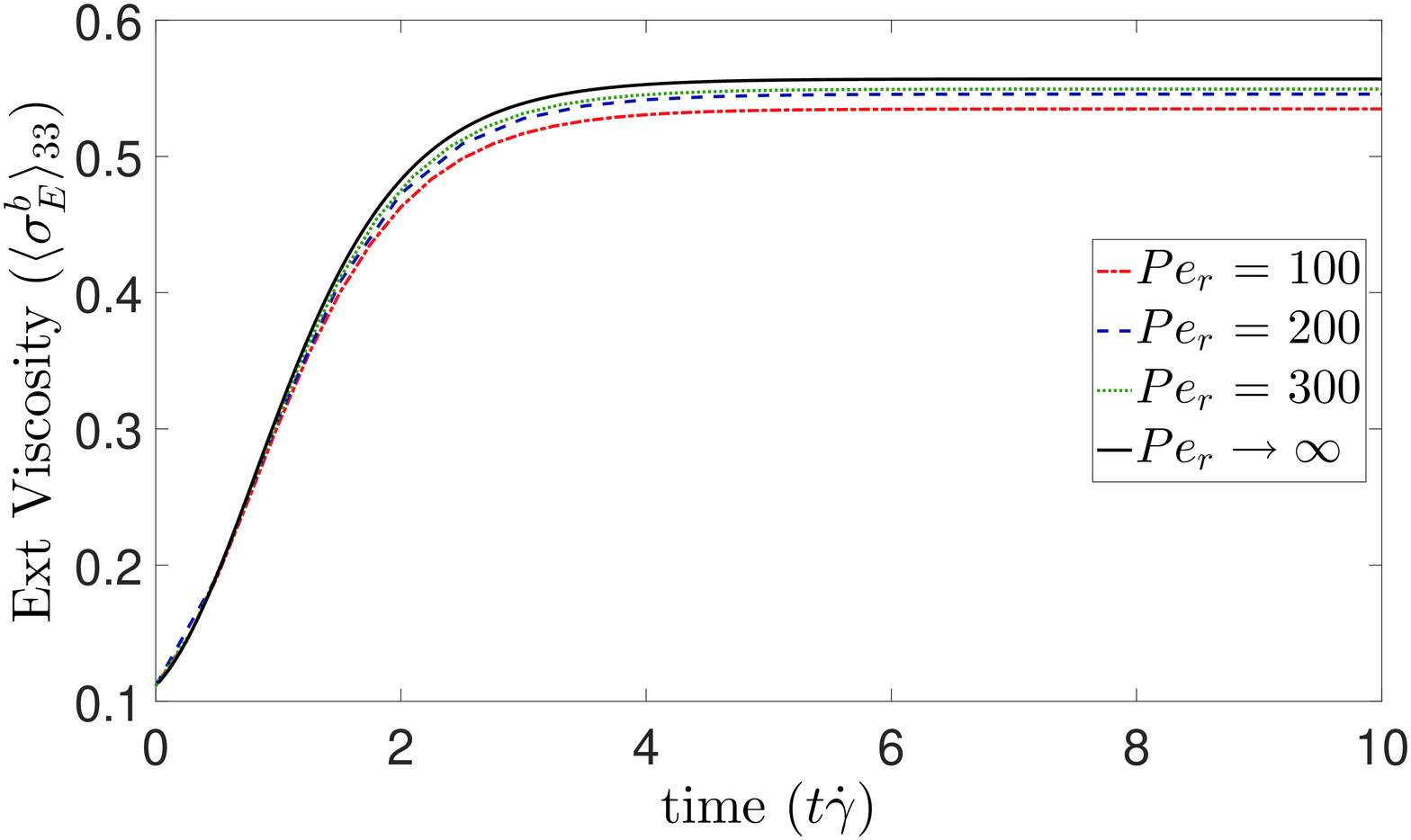}}
\subfigure[\hspace{-0.2cm}]{\includegraphics[scale=0.3]{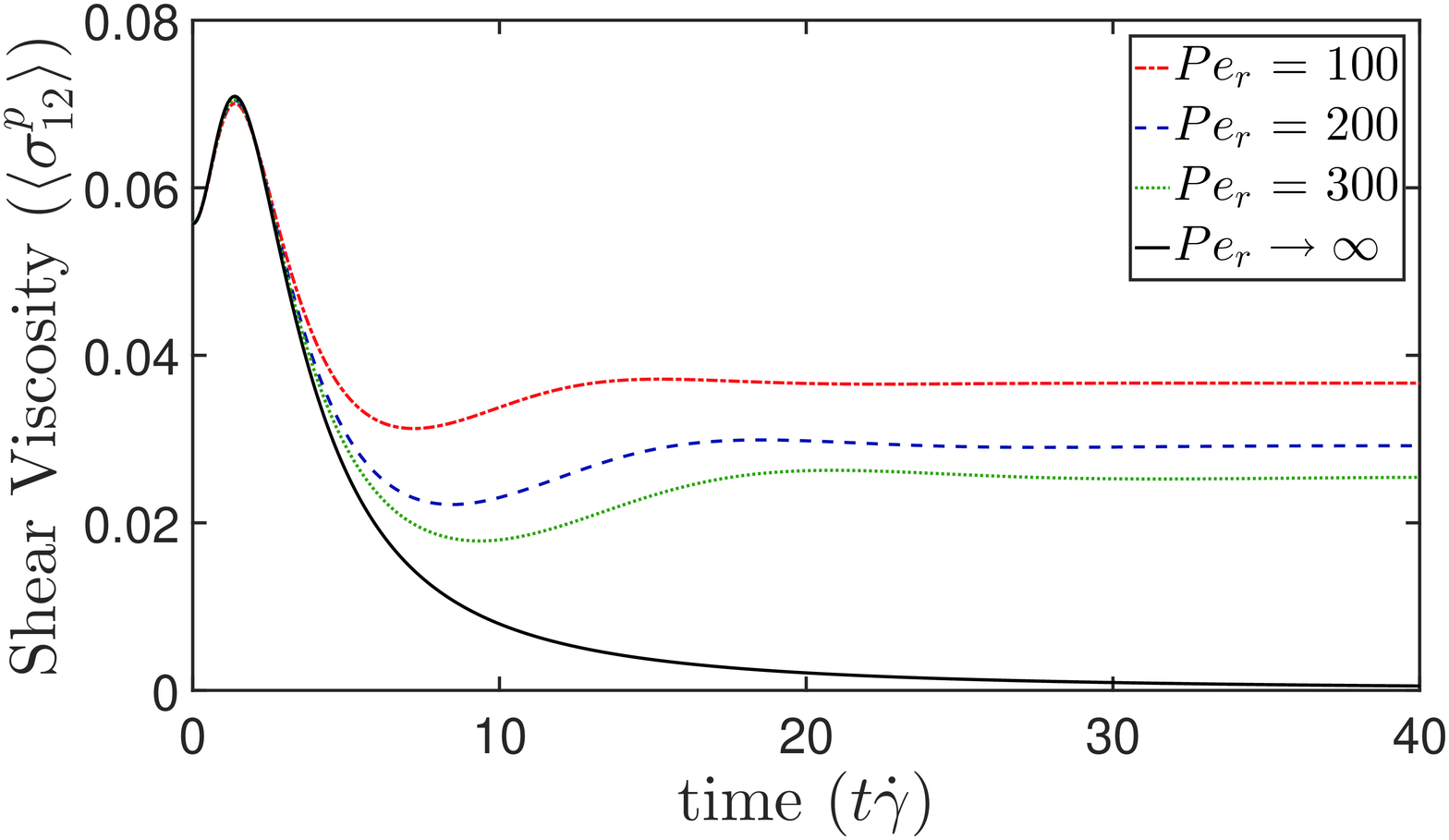}}
\end{center}
 \caption{Comparison of the numerical and analytical high-$\Pen$ relaxations. a) Axisymmetric extensional flow, b) simple shear flow at $\Pen_r = 100$, 200 and 300. The solid line represent the analytically obtained stress, (\ref{eq:extPassiveStress2}) and(\ref{eq:ssOmegaOuter3f}), respectively, while the dashed, dotted and dash-dotted lines represent the numerical solutions.}
 \label{high_pe_leading}
\end{figure}

We choose a slightly different coordinate system from that in \textsection\ref{subsubsec:ss}. In the analysis below, the polar angle $\theta$ is measured from the flow axis and the azimuthal angle $\phi$ from the gradient axis in the gradient-vorticity plane. The choice is motivated by the meridional nature of the infinite-aspect-ratio Jeffery trajectories. In the chosen coordinates, $\phi$ only acts to parametrize the trajectories, with only  $\theta$ varying along a given trajectory. Using $\dot{\boldsymbol{p}} = - \sin^2\theta\cos\phi\, \hat{\boldsymbol{\theta}}$, (\ref{eq:probNDextHigh}), now takes the form:
\begin{equation}
\frac{\partial\,\Omega_o}{\partial t} - \sin^2\theta\cos\phi\frac{\partial\,\Omega_o}{\partial \theta} - 3\sin\theta\cos\theta\cos\phi\,\Omega_o = 0.
 \label{eq:ssOmegaOuter1}
\end{equation}
Again, use of the method of characteristics gives:
\begin{subeqnarray}
  \frac{\mathrm{d} \theta}{\mathrm{d} t}  &=& \sin^2\theta\cos\phi,\\[3pt]
  \frac{\mathrm{d} \Omega_o}{\mathrm{d} t} &=& 3\sin\theta\cos\theta\cos\phi\Omega_o.
  \label{eq:ssOmegaOuter2}
\end{subeqnarray}
The solution to (\ref{eq:ssOmegaOuter2}a), gives a characteristic curve of the form $\cot\theta = t\cos\phi + \theta^\prime$, where $\theta^\prime$ is a constant of the integration. Using this  to solve (\ref{eq:ssOmegaOuter2}b), we get:
\begin{equation}
\Omega_o = \frac{1}{4\pi}\left(\frac{1}{1 + t^2\cos^2\phi\sin^2\theta - 2t\cos\phi\sin\theta\cos\theta}\right)^{\frac{3}{2}}.
 \label{eq:ssOmegaOuter3}
\end{equation}
Finally, the shear viscosity determined using (\ref{eq:ssOmegaOuter3}), is
\begin{equation}
\langle\sigma_{SS}^b\rangle_{12}^p = \frac{\pi(nL^3)}{6\ln\kappa}\int \mathrm{d}\boldsymbol{p} \sin^2\theta\cos^2\theta\cos^2\phi \,\Omega_o,
 \label{eq:ssOmegaOuter3f}
\end{equation}
the integral above being evaluated numerically. In figure \ref{high_pe_leading}b, we plot the passive viscosity obtained numerically in \textsection\ref{subsec:stress}, for $\Pen_r = 100$, 200 and 300, along with (\ref{eq:ssOmegaOuter3f}). Here, the initial peak in the viscosity is convective in origin, and as explained in \textsection\ref{sec:rheo}, is associated with the transient accumulation of swimmer orientations along the extensional axis on time scales of O($\dot{\gamma}^{-1}$). The viscosity predicted by (\ref{eq:ssOmegaOuter3f}) monotonically decays to zero thereafter. In contrast, the numerical solution exhibits an oscillatory approach towards a finite $\Pen$-dependent plateau. The oscillations are a signature of rotary diffusion becoming important at large times, as is evident from the lack of collapse, of the stress oscillations for different $\Pen$, on a convective time scale.

\section{Conclusion}\label{sec:conclusion}
In this paper, we have analysed the stress relaxation in an impulsively sheared bacterial suspension over the entire range of  the Peclet number $\Pen=\dot{\gamma}\tau$, where $\dot{\gamma}$ is the imposed shear rate and $\tau$ is the duration of a bacterial run. As shown in \cite{nambiar17}, the linear response is well accounted for by dilute suspension theory provided an additional elastic stress contribution due to activity, arising from the orientational anisotropy of the bacterium force dipoles, is accounted for in the suspension stress. The emphasis in this manuscript has been on the non-linear regime corresponding to Peclet numbers greater than about 0.2. We find a qualitative difference between our predictions of the stress relaxation in the non-linear regime and the experimental measurements of \cite{Gachelin2015}, particularly at the highest $\Pen$, as highlighted in figure \ref{exp_theo_comp}. The two main aspects of the disagreement are the magnitude of the steady state viscosity plateau at high $\Pen$, and the relaxation time required to reach the steady state (see \textsection\ref{sec:rheo}). The former is the more puzzling aspect of the disagreement, which appears to defy a reasonable explanation based on the response of a single bacterium to an imposed shear. In contrast to theory, the long-time viscosity at the highest values of $\Pen$ investigated in the experiments is higher than the initial stress jump. Within the theoretical formulation, the initial jump corresponds to the sum of the solvent stress and the instantaneous (viscous) hydrodynamic stress associated with an unoriented (isotropic) microstructure, while the plateau corresponds to the solvent stress plus the hydrodynamic stress for a nearly flow-aligned microstructure. At the highest $\Pen$, the active stress should be sub-dominant, and the aligned microstructure must therefore lead to a lower (passive) stress. The discrepancy in the relaxation times (the experimental relaxations being much slower at large $\Pen$) is perhaps of secondary importance, given the possible inability of the measuring device to resolve the extremely short time scales ($\sim 0.01$s) predicted for $\Pen = 64$.

Two contributions to the stress are not accounted for in this paper, which may possibly complicate the stress relaxation, but we believe that both are small, at least at large $\Pen$. The first is the active analog of the Brownian stress for passive suspensions, which has already been discussed in some detail in \cite{nambiar17}. As mentioned therein, the active stochastic stress contribution is in general expected to be different since the stochastic motion of the swimmer occurs under a torque-free constraint, and is therefore unlike a passive Brownian particle; the calculation of this stochastic stress contribution might be considerably involved owing to a dependence on the details of the swimming mechanism. The second, apparently novel, contribution to the bacterium stress is the so-called swim stress, identified recently by \cite{brady17}. This contribution appears to have an entropic origin, and is thought to be related to the flow-induced distortion of the persistent random motion of the swimmer. Since the stress originates at the level of the swimmer trajectory, and is not directly related to the swimmer force distribution, this swim viscosity contribution is predicted to be of the same sign for both pushers and pullers (unlike the active orientation stress included in the analysis here). A discussion of the detailed nature of this contribution is beyond the scope of the present manuscript, and we only note two points in this regard. The first is that a passive analog of the swim stress may also be identified for inertial Brownian suspensions, wherein particle inertia leads to a weak persistence in the Brownian motion of passive particles, in a manner similar to that associated with active Brownian particles (ABP's). The entropic stress that results from the flow-induced distortion of the thermal random walk is rarely discussed in the classical literature on account of its smallness, which in turn is due to the extremely small momentum relaxation time $\tau_p$ of a typical Brownian particle. Scaling arguments show that this passive swim stress is O($nk$T).($\dot{\gamma}\,\tau_p$), leading to a passive swim viscosity of O[$(nL^3) (kT \tau_p/L^3))$]; here, $kT$ is the thermal engery unit, and $\dot{\gamma}\tau_p$ is the particle Stokes number. The ratio of the swim to the Einstein viscosity comes out to be O($kT \rho_p/(\mu^2 L)$) which is negligibly small for typical Brownian particles. The second point, more relevant to this paper, is that the swim stress identified by \cite{brady17} is predicted to become vanishingly small at large $\Pen$, on account of the decay of the off-diagonal term of the diffusivity tensor.  Inclusion of the swim-stress relaxation cannot therefore account for the discrepancy between the theoretical and experimental high-$\Pen$ relaxations.

The analysis in this paper, and the dicussion above, are based on the presumption that the motion of bacteria are uncorrelated. However, it is widely recognized that a bacterial suspension transitions to a state of collective motion with increasing volume fraction or activity. Such a transition is likely to lead to differing stress relaxations on account of additional contributions to the stress arising from the work done by the ambient shear in orienting long-wavelength orientation and velocity fluctuations, that are characteristic of collective motion. The absolute magnitude of the viscosity is certainly expected to change across a critical volume fraction, on account of contributions arising from collective motion. Unfortunately, the experimental evidence with regard to stress transients, available from \cite{Gachelin2015}, is restricted to a single volume fraction ($\phi = 0.67\%$ or $nL^3 = 3.43$). There is additional information about these experiments in \cite{loisy2018active}, but this is only in the form of derived quantities, for instance, the volume fraction dependence of the effective viscosity. Any speculation with regard to the effect of collective motion on the stress relaxation would be premature in the absence of knowledge about the systematic variation in the observed relaxations with volume fraction.

Finally, recall that the preplexing aspect of the observed stress relaxation at high $\Pen$ is that of the long-time plateau being higher than the initial stress jump, the latter arising due to the viscous stress associated with inextensible isotropically oriented bacteria. If one assumes the bacteria to instead be deformable, then the instantaneous viscous stress will be replaced by an elastic stress that can only develop on a time scale of O($\mu/G$), $G$ being an effective elastic modulus. For sufficiently small $G$, the increasing elastic stress on time scales of O($\mu G$) might overwhelm the decrease in the orientation stress due to the bacterial alignment over a shorter time scale of O($\dot{\gamma}^{-1}$). In other words, for $G$ not too large, it is bacterium deformability and not orientability that would likely control the stress relaxation at high $\Pen$. The available force-extension data [\cite{darnton2007force}] for salmonella filaments suggests a force of the order of a few picoNewtons in order for the flagellar bundle to undergo a significant extension (this could be a flow-induced transition to a different polymorph). Assuming a spheroidal bacterium head of aspect ratio 2, with a diameter of O($1\mu m$), the viscous stress acting at $\dot{\gamma}\sim 60$ s$^{-1}$ (the highest value used in the experiments), leads to a hydrodynamic force that is only a fraction of a picoNewton in an aqueous ambient ($\mu \sim$ O($10^{-3}$) Pa-s), suggesting flagellar extension to perhaps only be of minor importance in the experiments of \cite{Gachelin2015}. However, \cite{darnton2007force} also found the flagellar bundle to buckle under the smallest compressive forces. Such a buckling therefore appears likely at high shear rates, in the time that the bacterium spends in the compressional quadrant. It is not clear as to what the signature of buckling on the stress transient would be. Earlier computations for non-Brownian passive flexible fibers [\cite{becker2001instability}] suggest the onset of normal stress differences as a rheological signature of a buckled microstructure. However, the authors also find buckling to lead to significant shear-thinning which would act to exacerbate the existing  discrepancy between the observed and computed high-$\Pen$ stress plateaus. The above discussion nevertheless highlights the many factors that could play a role in determining the stress relaxations at the higher shear rates. It would help to keep these in mind when designing future experiments on swimmer suspension rheology.

\begin{appendix}
\section{Higher order corrections for $\Pen\ll1$} \label{appendix_higherOrder}
 \subsection{Orientation distribution } \label{appendix_Omega}
 When $\Pen\ll1$, the effect of the imposed impulsive shear on the orientation distribution can be captured by a regular perturbation expansion about an isotropic base state. One writes $\Omega(\boldsymbol{p}, t)$ as [\cite{nambiar17}]:
 \begin{equation}
 \Omega(\boldsymbol{p}, t) = \Omega^{(0)} + \Pen\Omega^{(1)}(\boldsymbol{p}, t) + \Pen^2\Omega^{(2)}(\boldsymbol{p}, t) + \Pen^3\Omega^{(3)}(\boldsymbol{p}, t) + \ldots, 
  \label{eq:perturb}
 \end{equation}
where $\Omega^{(0)}=1/(4\pi)$.
 
Substituting (\ref{eq:perturb}) for $\Omega(\boldsymbol{p}, t)$ in the governing kinetic equation (\ref{eq:probND}) leads to the following sequence of time dependent equation at each order in $\Pen$:
\begin{equation}
\frac{\partial \Omega^{(n)}}{\partial t}  + \Big (\Omega^{(n)} - \int\!\! K({\boldsymbol{p}}|\boldsymbol{p^\prime}) \Omega^{(n)} (\boldsymbol{p^\prime}, t )\mathrm d\boldsymbol{p^\prime} \Big ) - (D_r\tau) \nabla^2_{\boldsymbol{p}} \Omega^{(n)} = - \bnabla_{\boldsymbol{p}} \!\cdot\! (\dot{\boldsymbol{p}} \Omega^{(n-1)}).
\label{eq:perturbationOrder}
\end{equation}

\cite{nambiar17} evaluated the Green's function of the operator on the left side of (\ref{eq:perturbationOrder}) given by:
\begin{eqnarray}
G(\boldsymbol{p}, t | \boldsymbol{p^\prime}, t^\prime) =  H(t - t^\prime)&&\mbox{}\left[\frac{1}{4\pi} + \sum_{n=1}^\infty \sum_{m=-n}^{n} Y_n^m(\boldsymbol{p}) Y_n^{m\ast}(\boldsymbol{p^\prime})\right.\nonumber\\&& \left.\mbox{}  \exp\left(\!\!-\! \big \{(1- \frac{4\pi A_n}{(2n \!+\! 1)}) + n(n\!+\!1)(\tau D_r)  \big\} (t\! -\! t^\prime) \right) \right]\astrut .  
 \label{eq:greens}
\end{eqnarray}

Note that the Green's function (\ref{eq:greens}) is dependent solely on the intrinsic dynamics of the swimmer, and therefore remains the same for both axisymmetric extension and simple shear. \cite{nambiar17} only determined the first effects of the anisotropy due to the imposed shear, as characterized by $\Omega^{(1)}$ below. Symmetry arguments suggest $\Omega^{(1)}\propto \boldsymbol{E}:\boldsymbol{pp}$, and \cite{nambiar17} showed that:
 \begin{equation}
 \Omega^{(1)} = \frac{3}{4\pi B_2} (\boldsymbol{E}:\boldsymbol{pp})f(t),
  \label{eq:firstOrder}
 \end{equation}
 where, $f(t)=H(t-T_1)(1-e^{-B_2(t-T_{1})})$, for a step initiation of the imposed flow. Here, $T_1$ is the time at which the step jump is initiated and $B_n = 1 + n(n+1)(D_r\tau) - 4\pi A_n/(2n+1)$; the $A_{n}$'s here are obtained from (\ref{eq:kernel}) by using the orthogonality of the Legendre polynomials:
 \begin{equation}
 A_n = \frac{\beta}{(4\pi\sinh\beta)}\frac{\int\exp(\beta\boldsymbol{p}\cdot\boldsymbol{p^\prime}) P_n(\boldsymbol{p}\cdot\boldsymbol{p^\prime}) \mathrm{d}(\boldsymbol{p}\cdot\boldsymbol{p^\prime})}{\int \left(P_n(\boldsymbol{p}\cdot\boldsymbol{p^\prime})\right)^2 \mathrm{d}(\boldsymbol{p}\cdot\boldsymbol{p^\prime})}
  \label{eq:An}
 \end{equation}
 
 Herein, we carry out the above expansion to higher orders in order to obtain the first effects of non-linearities that include the normal stress differences and the shear-rate dependence of the viscoisty. Such an expansion has been carried out earlier in the context of passive Brownian suspensions [\cite{brenner1974}].
 
 For axisymmetric extensional flow, at second order we obtain:
 \begin{eqnarray}
 \Omega^{(2)}_E = \frac{3}{4\pi B_2}&&\left[5 f^{(4)}(t) (\boldsymbol{E}:\boldsymbol{pp})^2 + \frac{1}{7}\left(-10f^{(4)}(t) + 3f^{(2)}(t)\right)(\boldsymbol{E}:\boldsymbol{pp})\right.\nonumber\\&&\left.\mbox{} - \frac{2}{3} f^{(4)}(t)\left(\boldsymbol{E}:\boldsymbol{E}\right)\right].
  \label{eq:2ndOrderE}
 \end{eqnarray}

 For simple shear flow, at second and third order, we have:
 \begin{subeqnarray}
  \Omega^{(2)}_{S}  &=& \frac{3}{4\pi B_2} 
 \mbox{} \left[5 f^{(4)}(t) (\boldsymbol{E}:\boldsymbol{pp})^2 - \frac{1}{7}\left(20 f^{(4)}(t)-6 f^{(2)}(t)\right) (\boldsymbol{E}\cdot\boldsymbol{p})\cdot(\boldsymbol{E}\cdot\boldsymbol{p})\right.\nonumber\\&&\left. \mbox{} - 2f^{(2)}(t)(\boldsymbol{\omega}\cdot\boldsymbol{p})\cdot(\boldsymbol{E}\cdot\boldsymbol{p})+\frac{2}{7}\left(f^{(4)}(t) - f^{(2)}(t)\right)(\boldsymbol{E}:\boldsymbol{E})\right ]  \astrut   ,\\
  \Omega^{(3)}_{S}  &=& \frac{3}{4\pi B_2} 
 \mbox{} \left[-35 f^{(46)}(t) (\boldsymbol{E}:\boldsymbol{pp})^3  \right.\nonumber\\ && \left. \mbox{}+\left(\frac{420}{11} f^{(46)}(t)-\frac{300}{77} f^{(44)}(t)-\frac{30}{7}f^{(24)}(t)  \right)(\boldsymbol{E}:\boldsymbol{pp})(\boldsymbol{E}\cdot\boldsymbol{p})\cdot(\boldsymbol{E}\cdot\boldsymbol{p})\right.\nonumber\\&&\left.\mbox{} +10\left(2 f^{(44)}(t)+ f^{(24)}(t)  \right)(\boldsymbol{E}:\boldsymbol{pp})(\boldsymbol{\omega}\cdot\boldsymbol{p})\cdot(\boldsymbol{E}\cdot\boldsymbol{p}) \right.\nonumber\\ &&\left. \mbox{} + \left(-\frac{35}{11} f^{(46)}(t)+\frac{450}{539} f^{(44)}(t)+\frac{24}{49}f^{(42)}(t) + \frac{45}{49}f^{(24)}(t) + \frac{46}{49}f^{(22)}(t)\right)(\boldsymbol{E}:\boldsymbol{pp})\right ].  \nonumber\\&&\left.\mbox{}\astrut \right. 
    \label{eq:2ndOrderSS}
 \end{subeqnarray}

 The functions of time in (\ref{eq:2ndOrderE}) and (\ref{eq:2ndOrderSS}), i.e. $f^{(n)}$'s and $f^{(ij)}$'s are as follows:
 \begin{subeqnarray}
  f^{(n)}(t)  &=& \int_0^t \mathrm{d} t^\prime H(t-t^\prime) \dot{\gamma}(t^\prime) f(t^\prime)\exp\left(-B_n(t-t^\prime)\right), \\
  f^{(ij)}(t)  &=& \int_0^t \mathrm{d} t^\prime H(t-t^\prime)\dot{\gamma}(t^\prime) f^{(i)}(t^\prime)\exp\left(-B_j(t-t^\prime)\right),
    \label{eq:timeFunctions}
 \end{subeqnarray}
where $\dot{\gamma}(t) = H(t-T1)$.

 Using (\ref{eq:2ndOrderE}) - (\ref{eq:timeFunctions}), we now proceed to evaluate the extensional viscosity, the two normal stress differences and the shear viscosity. 

 \subsection{Expressions for normal stress differences and viscosity} \label{appendix_Stress}
 At any given order in $\Pen$, on substituting for $\Omega(\boldsymbol{p}, t)$ using (\ref{eq:perturb}) in (\ref{eq:bactstress3}), we have:
 \begin{eqnarray}
  \langle\boldsymbol{\sigma}^b\rangle^{(n)} = && \Pen^{n-1}\left[ \frac{\pi (nL^{3})}{6 \ln \kappa}\dot{\gamma}(t)\int \mathrm{d}\boldsymbol{p} \, \Omega^{(n-1)}(\boldsymbol{p},t)(\boldsymbol{E}:\boldsymbol{pp})\left(\boldsymbol{pp}-\frac{\boldsymbol{I}}{3}\right)\right. \nonumber\\  &&\left.\mbox{} - \frac{\alpha}{2M}(nUL^2\tau)\int \mathrm{d}\boldsymbol{p} \, \Omega^{(n)}(\boldsymbol{p},t) \left(\boldsymbol{pp}-\frac{\boldsymbol{I}}{3}\right)\right]\astrut,
\label{eq:bactstress_Appendix}
\end{eqnarray}
 where we note that the passive contribution to the stress involves the term in the series for $\Omega(\boldsymbol{p}, t)$, which is one order less than the correction order being considered. The stress tensor in (\ref{eq:bactstress_Appendix}) has been non-dimensionalized by $\mu\dot{\gamma}$.
 
 We begin here by reiterating the results of \cite{nambiar17} for the leading order stress which is same for both axisymmetric extension and simple shear and is given as:
   \begin{equation}
  \langle\boldsymbol{\sigma}^b\rangle^{(1)} =  \frac{1}{B_{2}} \left[ \frac{ \pi B_2 (nL^{3})}{45 \ln \kappa}\dot{\gamma}(t) \boldsymbol{E}  - \frac{2}{5}\left(nUL^2\tau\right)\left(\frac{\alpha}{2M}\right)f(t)\boldsymbol{E}\right].
 \label{eq:bactstress_Appendix_Leading}
 \end{equation}
 
 We now proceed to higher orders. For axisymmetric extensional flow, substituting (\ref{eq:firstOrder}) and (\ref{eq:2ndOrderE}) in (\ref{eq:bactstress_Appendix}), gives:
  \begin{equation}
  \langle\boldsymbol{\sigma}_{E}^b\rangle^{(2)} =  \frac{\Pen}{B_{2}} \left[ \frac{2 \pi (nL^{3})}{105 \ln \kappa}\dot{\gamma}(t)f(t) \boldsymbol{E}  - \frac{6}{35}\left(nUL^2\tau\right)\left(\frac{\alpha}{2M}\right)f^{(2)}(t)\boldsymbol{E}\right],
 \label{eq:bactstress_Appendix_Ext}
 \end{equation}
 and for simple shear flow, substituting (\ref{eq:firstOrder}) and (\ref{eq:2ndOrderSS}) in (\ref{eq:bactstress_Appendix}), gives:

  \begin{subeqnarray}
  \langle\boldsymbol{\sigma}_{S}^b\rangle^{(2)}  &=& \frac{\Pen}{B_{2}} 
 \mbox{} \left[\frac{4\pi (nL^{3})}{105 \ln \kappa}\dot{\gamma}(t)f(t)\left(\boldsymbol{E}\cdot\boldsymbol{E}-(\boldsymbol{E}:\boldsymbol{E})\frac{\boldsymbol{I}}{3}\right) \right.\nonumber\\&&\left. \mbox{} - \frac{6}{5}\left(nUL^2\tau\right)\left(\frac{\alpha}{2M}\right)f^{(2)}(t)\left\{\frac{1}{3}\left(\boldsymbol{E}\cdot\boldsymbol{\omega}-\boldsymbol{\omega}\cdot\boldsymbol{E}\right)\right.\right.\nonumber\\ && \left.\left. \mbox{}- \frac{2}{7}\left(\boldsymbol{E}\cdot\boldsymbol{E}-(\boldsymbol{E}:\boldsymbol{E})\frac{\boldsymbol{I}}{3}\right)\right\}\right ]  \astrut ,\\
  \langle\boldsymbol{\sigma}_{S}^b\rangle^{(3)}  &=& \frac{\Pen^2}{B_{2}} 
 \mbox{} \left[-\frac{\pi (nL^{3})}{\ln \kappa}\dot{\gamma}(t)\left(\frac{4}{245}f_4(t) +\frac{2}{735} f_2(t)\right)\boldsymbol{E} \right.\nonumber\\&&\left. \mbox{} - 3\left(nUL^2\tau\right)\left(\frac{\alpha}{2M}\right)\left(\frac{92}{735}f^{(22)}(t) + \frac{16}{245}f^{(42)}(t)\right) \boldsymbol{E}\right ].  \astrut 
    \label{eq:bactstress_Appendix_SS}
 \end{subeqnarray}
 
  It is of interest to note that for axisymmetric extension, both first and second order stresses are proportional to $\boldsymbol{E}$. However, this is not the case with simple shear. The second order correction leads to only diagonal terms in the stress tensor, allowing one to capture the normal stress differences. The shear rate dependence of the viscosity is obtained at the third order.
 
 Using (\ref{eq:bactstress_Appendix_SS}$a$), we obtain expressions for the normal stress differences, $N_1 = (\langle\sigma_S^b\rangle_{11} - \langle\sigma_S^b\rangle_{22})/(\dot{\gamma}\tau)$ ($\equiv (\sigma_{xx} - \sigma_{yy})/(\mu\dot{\gamma}^2\tau)$) and $N_2 = (\langle\sigma_S^b\rangle_{22} - \langle\sigma_S^b\rangle_{33})/(\dot{\gamma}\tau)$ ($\equiv (\sigma_{yy} - \sigma_{zz})/(\mu\dot{\gamma}^2\tau)$):
  \begin{subeqnarray}
  N_1  &=& -\frac{2}{5 B_2}\left(\frac{\alpha}{2 M}\right)(nUL^2\tau) f_2(t) ,\\
   N_2  &=& \frac{\tau}{35 B_2}\left[\frac{\pi(nL^3)}{3 \ln\kappa}\dot{\gamma}(t)f(t) + 4 \left(\frac{\alpha}{2 M}\right) (nUL^2\tau) f_2(t)\right].  \astrut 
    \label{eq:bactstress_Appendix_NSD}
  \end{subeqnarray}
  
  The extensional and shear viscosities, $\langle\sigma^b_{E}\rangle_{33}$ and $\langle\sigma^b_{S}\rangle_{12}$ (note that this is the same as $\langle\sigma^b\rangle_{13}$ in \textsection\ref{subsubsec:ss2}, on account of the different coordinate system adopted), respectively, can be obtained from (\ref{eq:bactstress_Appendix_Leading}), (\ref{eq:bactstress_Appendix_Ext}) and (\ref{eq:bactstress_Appendix_SS}$b$).
  
 \begin{eqnarray}
  \langle\sigma_E^b\rangle_{33} =  \frac{1}{B_{2}} &&\left[ \frac{ \pi (nL^{3})}{15 \ln \kappa}\dot{\gamma}(t)\left(\frac{B_2}{3} + \frac{2}{7}\Pen f(t)\right)\right.\nonumber\\ && \left. \mbox{} - \frac{2}{5}\left(nUL^2\tau\right)\left(\frac{\alpha}{2M}\right)\left(f(t) + \frac{3}{7}\Pen f^{(2)}(t)\right)\right],
 \label{eq:viscosity_Appendix_E}
 \end{eqnarray}
 \begin{eqnarray}
  \langle\sigma_S^b\rangle_{12} =  \frac{1}{B_{2}} &&\left[ \frac{ \pi (nL^{3})}{2 \ln \kappa}\dot{\gamma}(t)\left(\frac{B_2}{45} + \Pen^2 \{f_4(t) - f_2(t)\}\right)\right.\nonumber\\ && \left. \mbox{} - \frac{1}{5}\left(nUL^2\tau\right)\left(\frac{\alpha}{2M}\right)\left(f(t) + \Pen^2 \left\{\frac{138}{147}f^{(22)}(t) + \frac{24}{49}f^{(42)}(t)\right\}\right)\right].\nonumber\\ &&
 \label{eq:viscosity_Appendix_S}
 \end{eqnarray} 
  
 \section{Validation of numerical scheme} \label{appendix_validation}
 
\begin{figure}
\begin{center}
\includegraphics[scale=0.3]{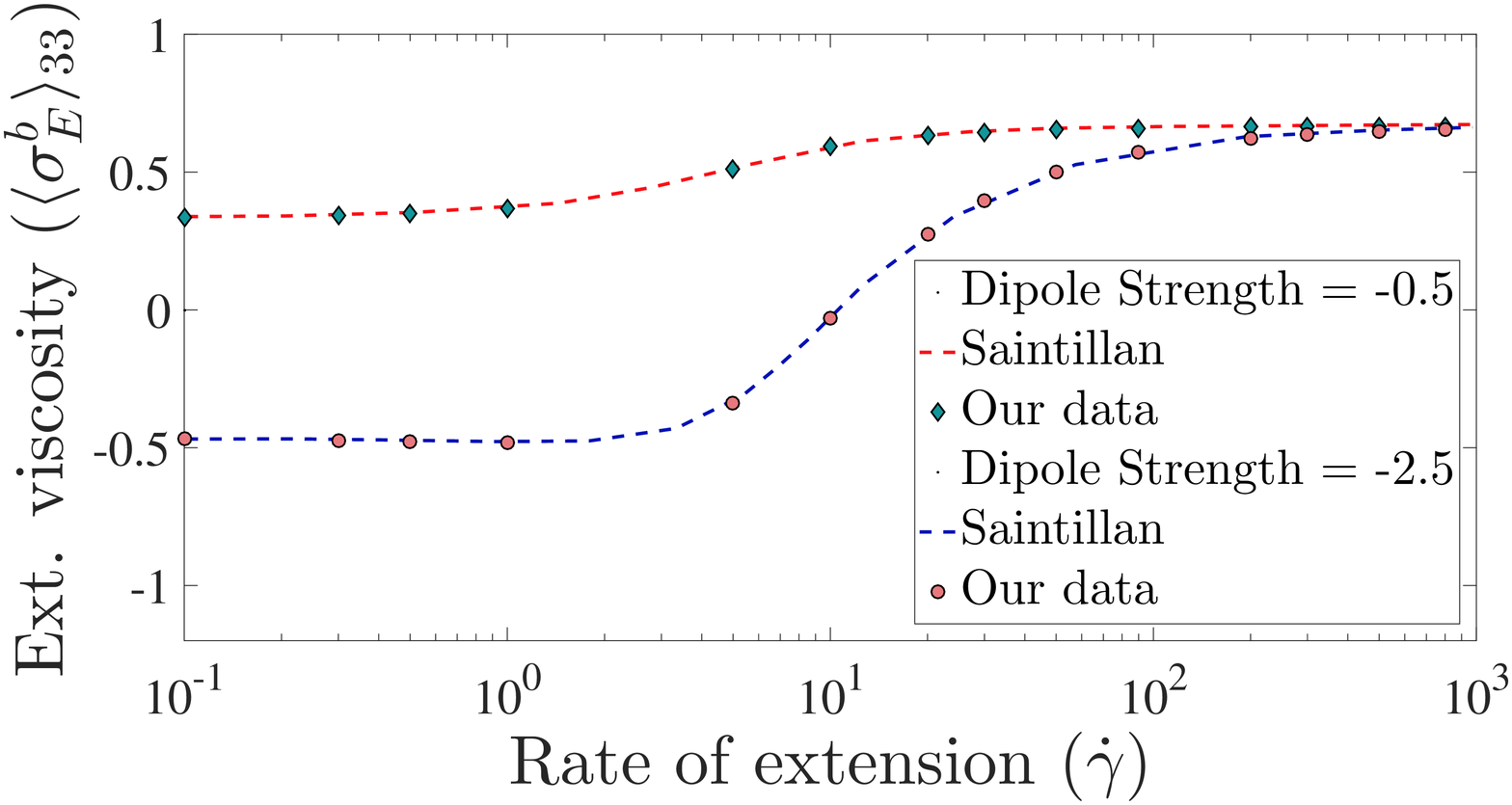}
\end{center}
 \caption{Comparison of steady-state extensional viscosity with \cite{saintillan2010extensional} for two values of dipole strength that were chosen by them. The dashed lines represent data obtained from \cite{saintillan2010extensional}, while the markers ($\Diamond$ and $\circ$) represent our data.}
 \label{saintillan_comparison1}
\end{figure}

\begin{figure}
\begin{center}
\subfigure[\hspace{-0.2cm}]{\includegraphics[scale=0.3]{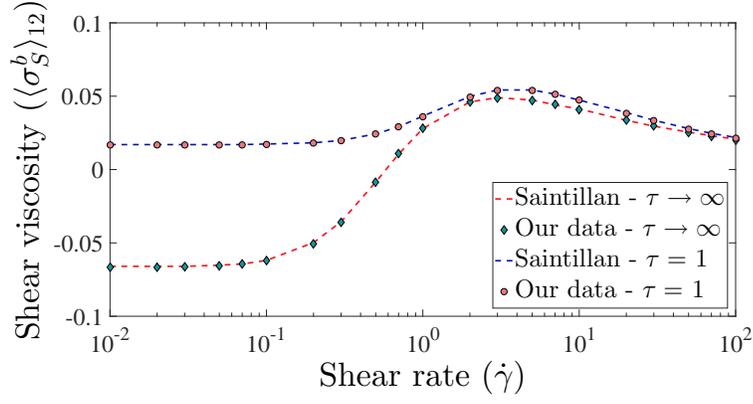}}
\subfigure[\hspace{-0.2cm}]{\includegraphics[scale=0.3]{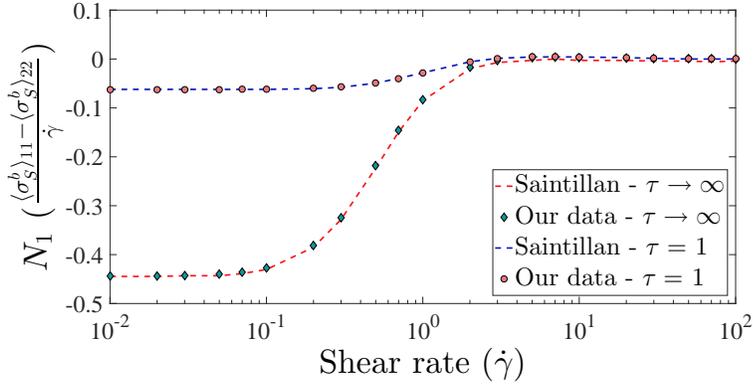}}
\subfigure[\hspace{-0.2cm}]{\includegraphics[scale=0.3]{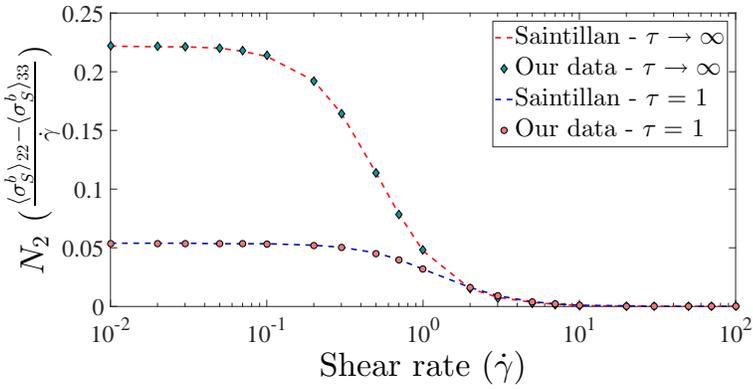}}
\end{center}
 \caption{Comparison of steady-state rheological properties with \cite{Saintillan2010}: a) Shear viscosity, b) First normal stress difference ($N_1$) and c) Second normal stress difference ($N_2$). The dashed lines represent data obtained from \cite{Saintillan2010}, while the markers ($\Diamond$, $\circ$) represent our data.}
 \label{saintillan_comparison2}
\end{figure}

\begin{figure}
\begin{center}
\subfigure[\hspace{-0.2cm}]{\includegraphics[scale=0.3]{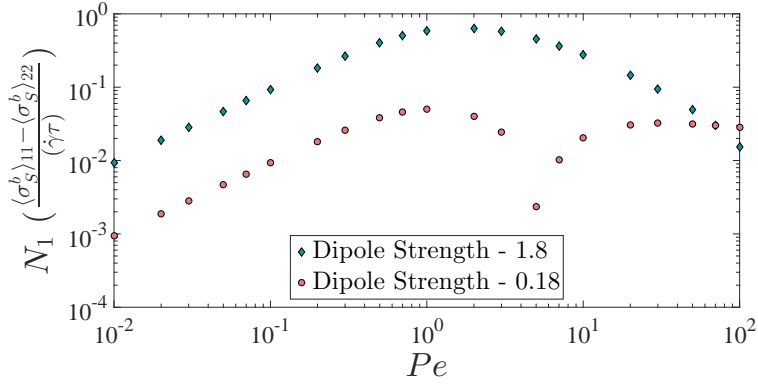}}
\subfigure[\hspace{-0.2cm}]{\includegraphics[scale=0.3]{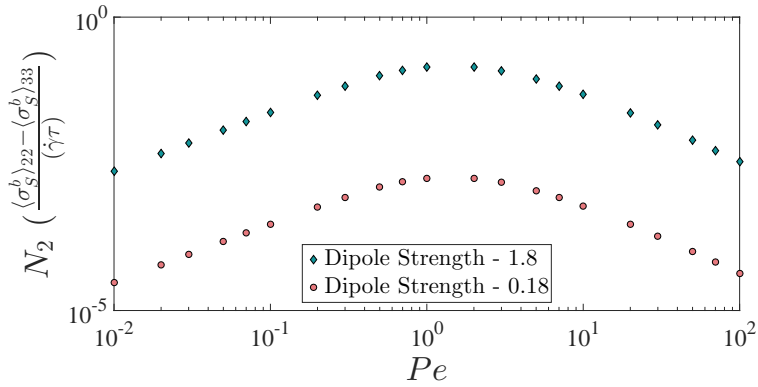}}
\end{center}
 \caption{Comparison of absolute values of the steady-state a) First normal stress difference ($N_1$) and b) Second normal stress difference ($N_2$), for varying dipole strength. The $\Diamond$ marker correspond to the dipole strength we consider in the main text (and this is representative of the experiments of \cite{Gachelin2015}) and the $\circ$ marker corresponds to 1/$10^{th}$ of the same. The discontinuity in (a) for the reduced dipole strength ($\circ$ marker) is indicative of a sign change in $N_1$}
 \label{saintillan_comparison3}
\end{figure}

In order to validate our numerical scheme, we first compare the steady-state properties with those that have been reported earlier [\cite{Saintillan2010, saintillan2010extensional}]. Note that the governing equation for the orientation probability density (\ref{eq:probND}) under steady-state conditions is identical to that used by \cite{Saintillan2010, saintillan2010extensional}; the only difference being that tumbling was absent for extensional flow in the Saintillan analysis. In figure \ref{saintillan_comparison1}, the steady extensional viscosity obtained from our numerical calculations is compared with figure 2 of \cite{saintillan2010extensional} for two values of the dipole strength chosen by them ($\tau D_r = 0$). In figure \ref{saintillan_comparison2}, the shear viscosity, first and second normal stress differences obtained from numerical calculations are compared for $\tau D_r = 0.1$ and $\tau D_r\rightarrow\infty$. In both the cases our analysis is in good agreement with that of \cite{Saintillan2010, saintillan2010extensional}. Figure \ref{saintillan_comparison3} highlights the scaling of the normal stress differences with $\Pen$. At low $\Pen$ both $N_1$ and $N_2$ scale as O($\Pen$), and is consistent with the O($\Pen^2$) analysis in \textsection\ref{appendix_Stress}.
 
\begin{figure}
\begin{center}
\subfigure[\hspace{-0.2cm}]{\includegraphics[scale=0.3]{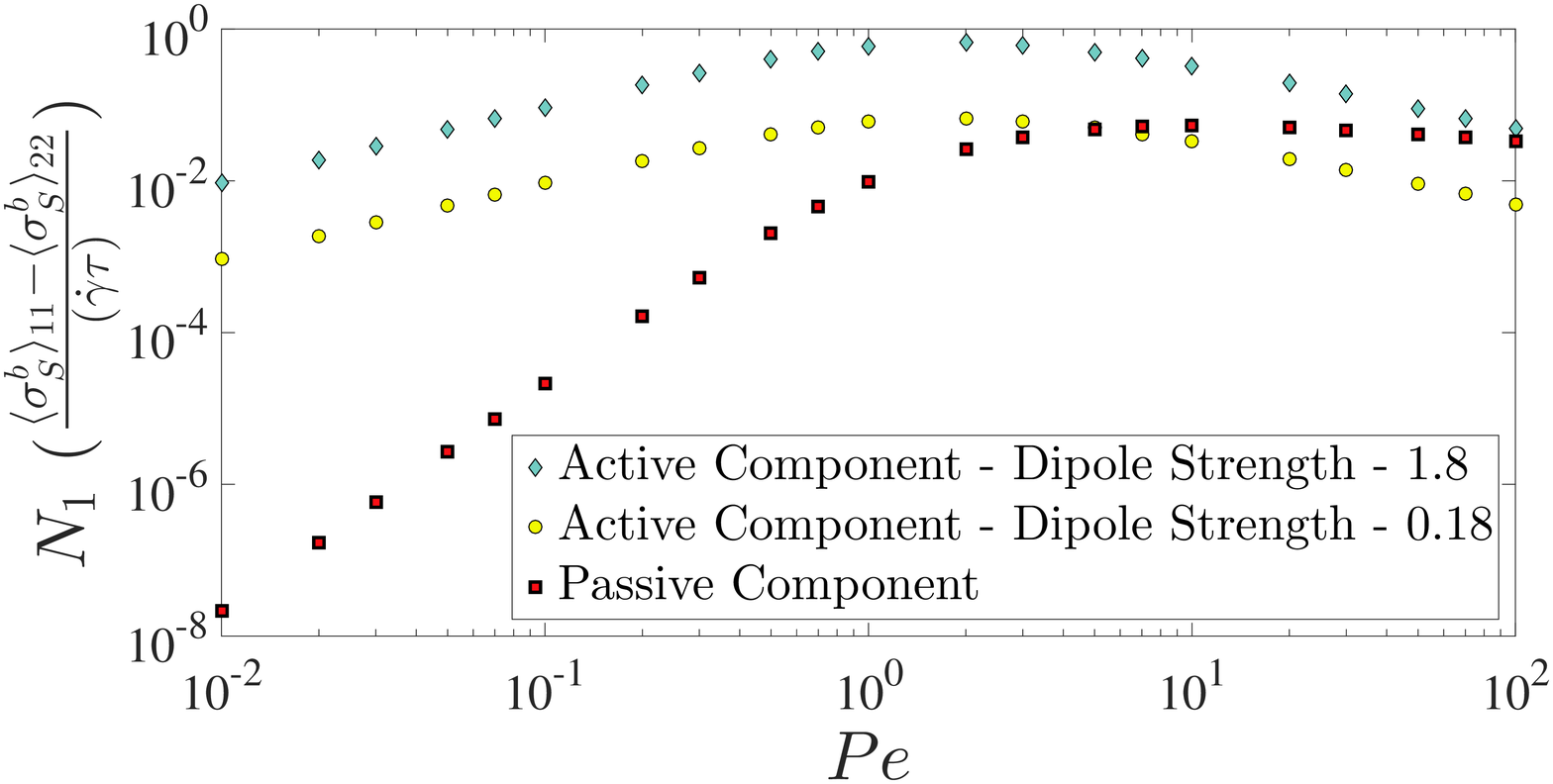}}
\subfigure[\hspace{-0.2cm}]{\includegraphics[scale=0.3]{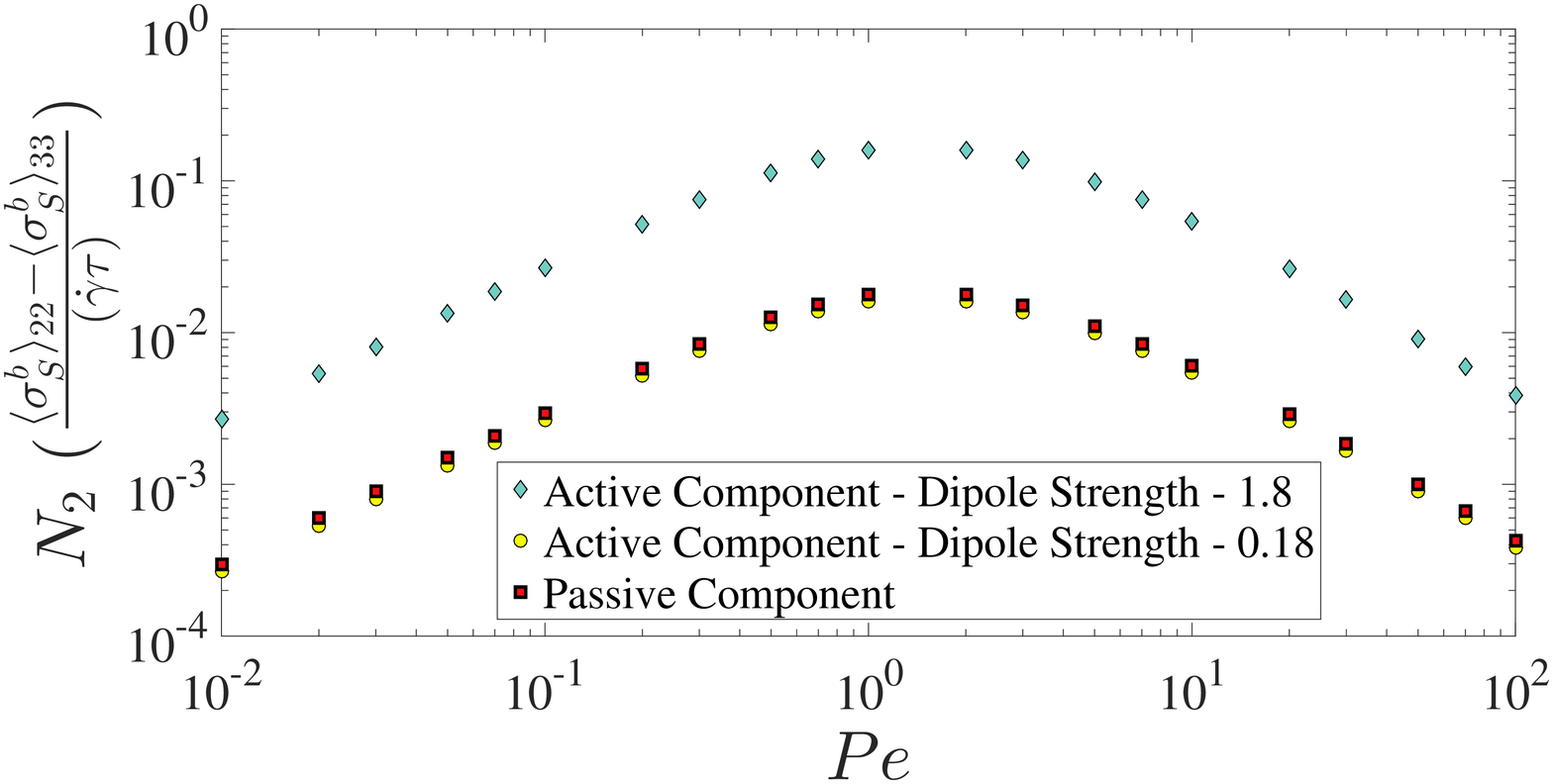}}
\end{center}
 \caption{Comparison of the absolute values of the active and passive contributions at steady-state a) First normal stress difference ($N_1$) and b) Second normal stress difference ($N_2$), for varying dipole strength. The $\Diamond$ and $\circ$ markers correspond to the contribution from the active component for the respective dipole strengths of 1.8 and 0.18. The $\square$ markers corresponds to the contribution from the passive component.}
 \label{saintillan_comparison4}
\end{figure}

When $\Pen\gg 1$, the scaling of both the NSD is O($\Pen^{-4/3}$) for the values of the swimmer parameters considered in the main text. In this regime, the passive contribution to the stress is dominant, and therefore one would expect the NSD to exhibit a response similar to that known for passive suspensions [\cite{strand1987, brenner1974}]. A change in sign (representing a transition from active to passive dominance) should be observed, with $\Pen$ corresponding to this sign change being dependent on the non-dimensional dipole strength of the swimmer, $\alpha/(2M)$. This large $\Pen$ transition appears rather subtle. For the swimmer parameters considered in the main text, the steady-state NSD do not undergo any sign change. In figure \ref{saintillan_comparison3}a, it is shown that at $\Pen = 5$, $N_1$ undergoes a change in sign (negative to positive), with the scaling transitioning from being O($\Pen^{-4/3}$) to O($\Pen^{-1/3}$), when the dipole strength is a tenth of what we have considered in the main text. At still smaller values of the dipole strength, the transition occurs earlier. For $N_2$ we do not observe any sign change with increasing $\Pen$ (figure \ref{saintillan_comparison3}b). In figures \ref{saintillan_comparison4}a and b, we compare the relative magnitudes of the active and passive contributions to the NSD. From these, it is evident that the absence of a crossover from the active to the passive regime for $N_2$ is due to the fact that both the active and passive contributions follow a $\Pen^{-4/3}$ scaling at large-$\Pen$. Thus, with increasing (non-dimensional) dipole strength, the response goes from being entirely dictated by active stress to being dictated by passive stress.

\begin{figure}
\begin{center}
\subfigure[\hspace{-0.2cm}]{\includegraphics[scale=0.3]{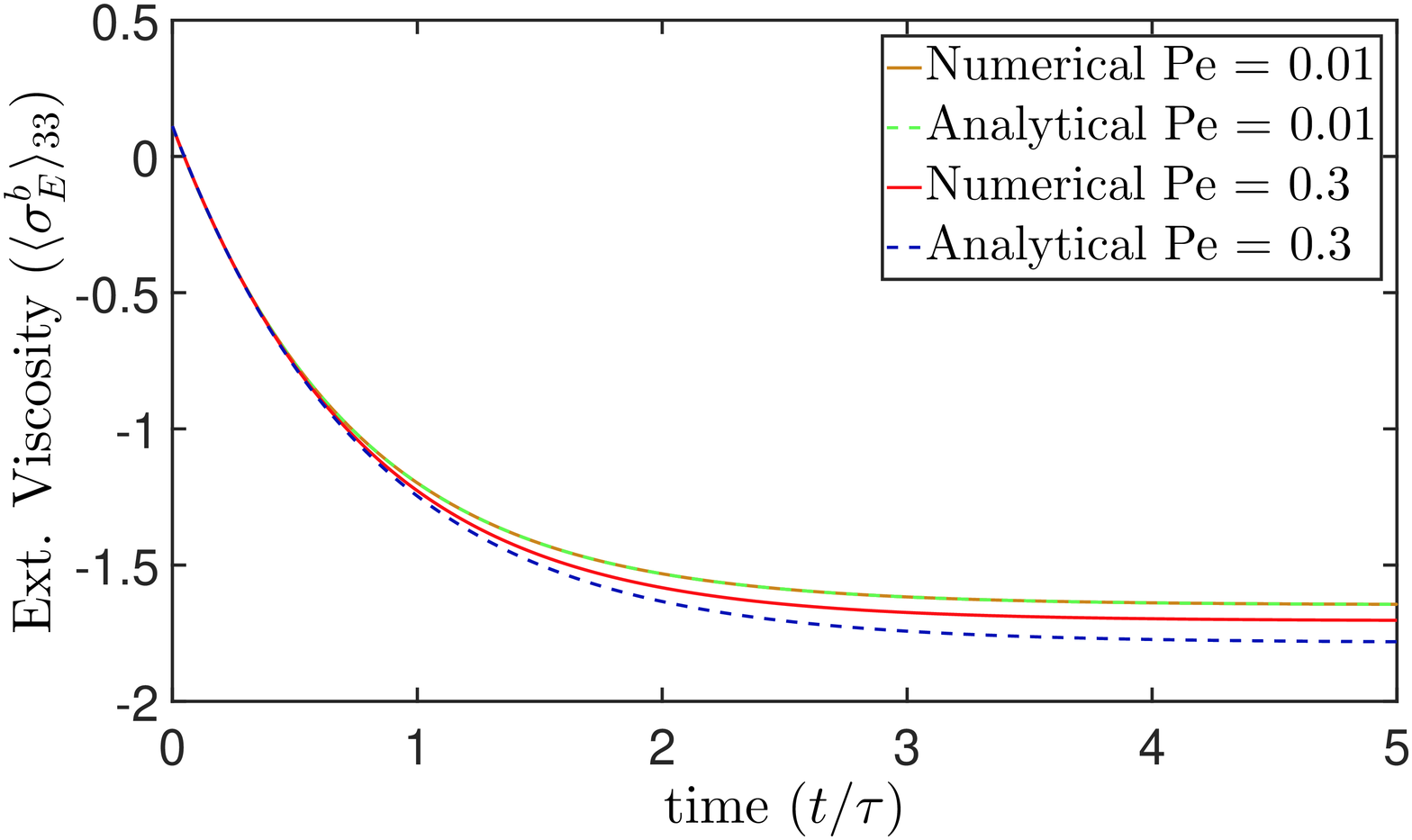}}
\subfigure[\hspace{-0.2cm}]{\includegraphics[scale=0.3]{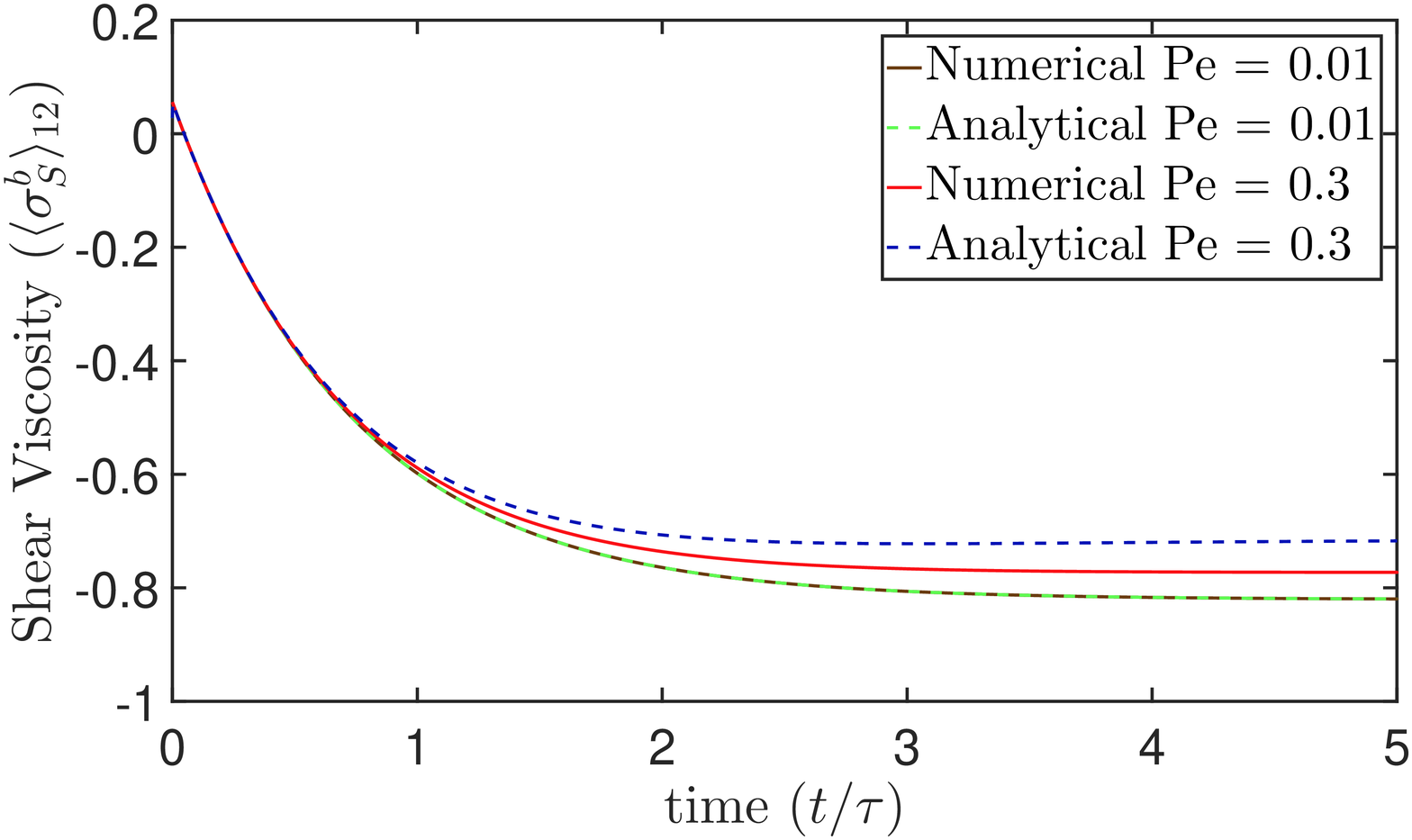}}
\end{center}
 \caption{Comparison of time dependent responses of bacterial viscosities at low $\Pen$ obtained numerically with the a) Extensional viscosity, with the analytical results evaluated to O($\Pen^2$), and b) Shear viscosity, with the analytical results evaluated to O($\Pen^3$). The dashed lines represent the analytically obtained viscosities, while the solid lines correspond to the viscosities obtained numerically.}
 \label{lowPe_comparison_visc}
\end{figure}

\begin{figure}
\begin{center}
\subfigure[\hspace{-0.2cm}]{\includegraphics[scale=0.3]{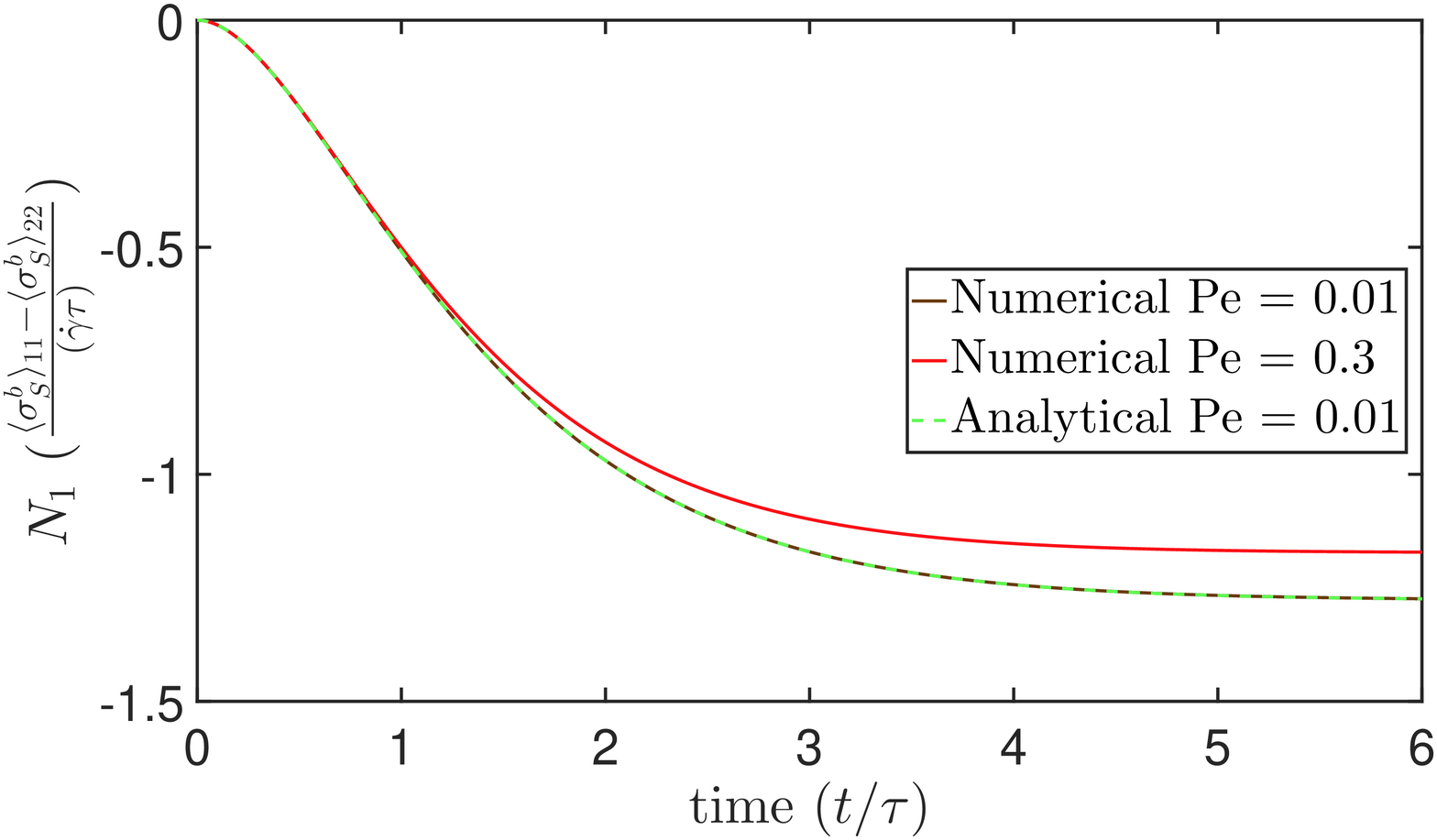}}
\subfigure[\hspace{-0.2cm}]{\includegraphics[scale=0.3]{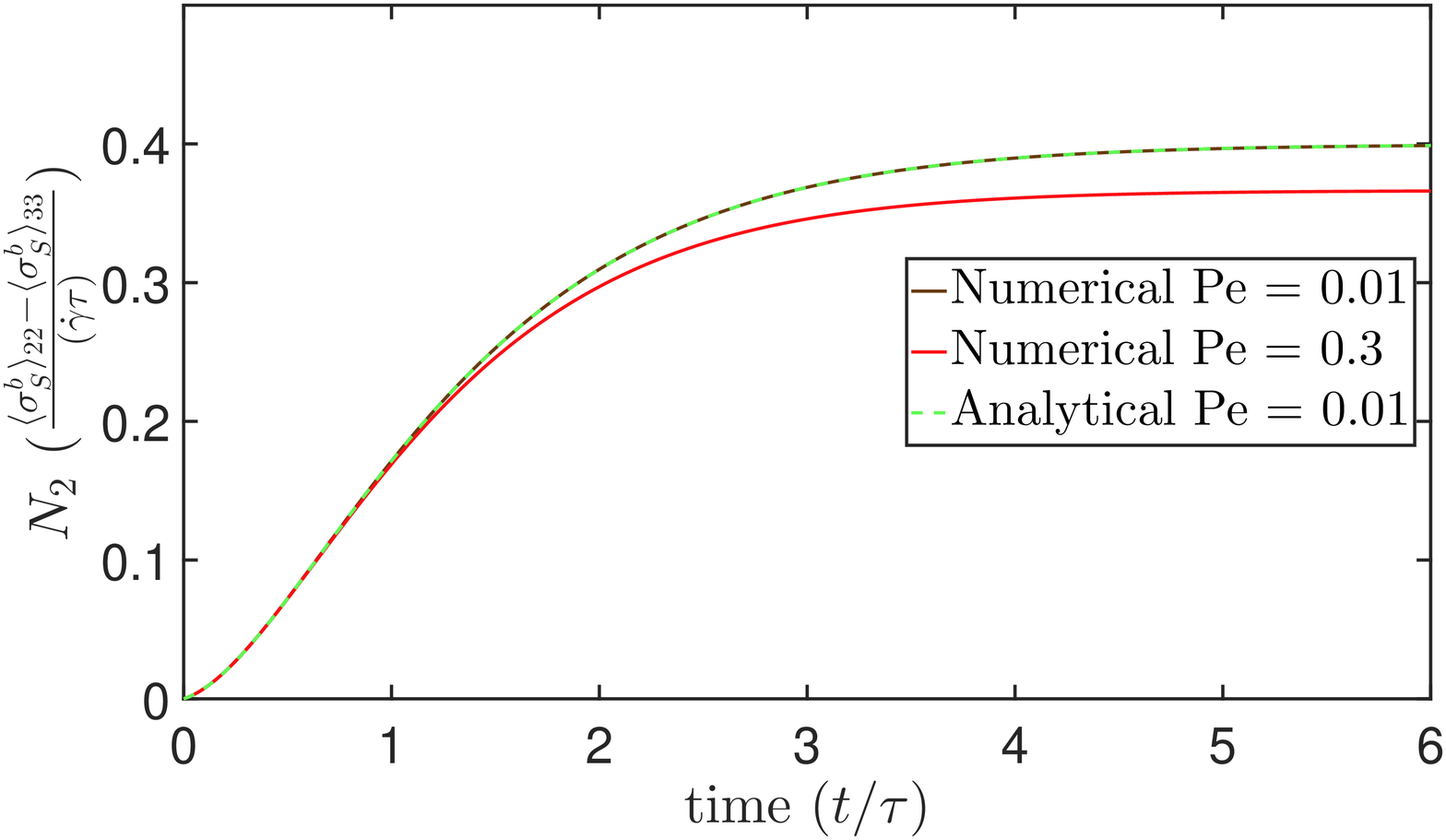}}
\end{center}
 \caption{Comparison of time dependent responses of bacterial normal stress differences at low $\Pen$ obtained from the O($\Pen^2$) analysis with numerics for the case of simple shear flow. a) $N_{1}$ b) $N_{2}$. The dashed lines represent the analytically obtained viscosities, while the solid lines correspond to the viscosities obtained numerically.}
 \label{lowPe_comparison_nsd}
\end{figure}

Next, the time dependent rheological porperties derived in \textsection\ref{appendix_Stress} are compared with those in \textsection \ref{subsec:stress}. In figures \ref{lowPe_comparison_visc}a and b, we show the extensional and shear viscosities, respectively, for two values of $\Pen$, viz., $\Pen =$ 0.01 and $\Pen =$ 0.3. In both cases, the theoretical and numerical curves are coincident for $\Pen = 0.01$, whereas the relative error at $\Pen = 0.3$ is about 7.4$\%$ for simple shear and 4.6$\%$ for unaxial extension, at steady state. In figure \ref{lowPe_comparison_nsd}, we plot the normal stress differences (NSD) for the same $\Pen$'s. At $\Pen=0.01$, the O($\Pen^2$) predictions for $N_1$ and $N_2$ match well with those determined numerically. For $\Pen = 0.3$, there is a mismatch at steady state of about 9$\%$. Note that accounting for the shear rate dependence of the normal stress differences  requires going to O($\Pen^4$), and the small-$\Pen$ theory in the earlier section is limited to O($\Pen^3$).

\begin{figure}
\begin{center}
\subfigure[\hspace{-0.8cm}]{\includegraphics[scale=0.3]{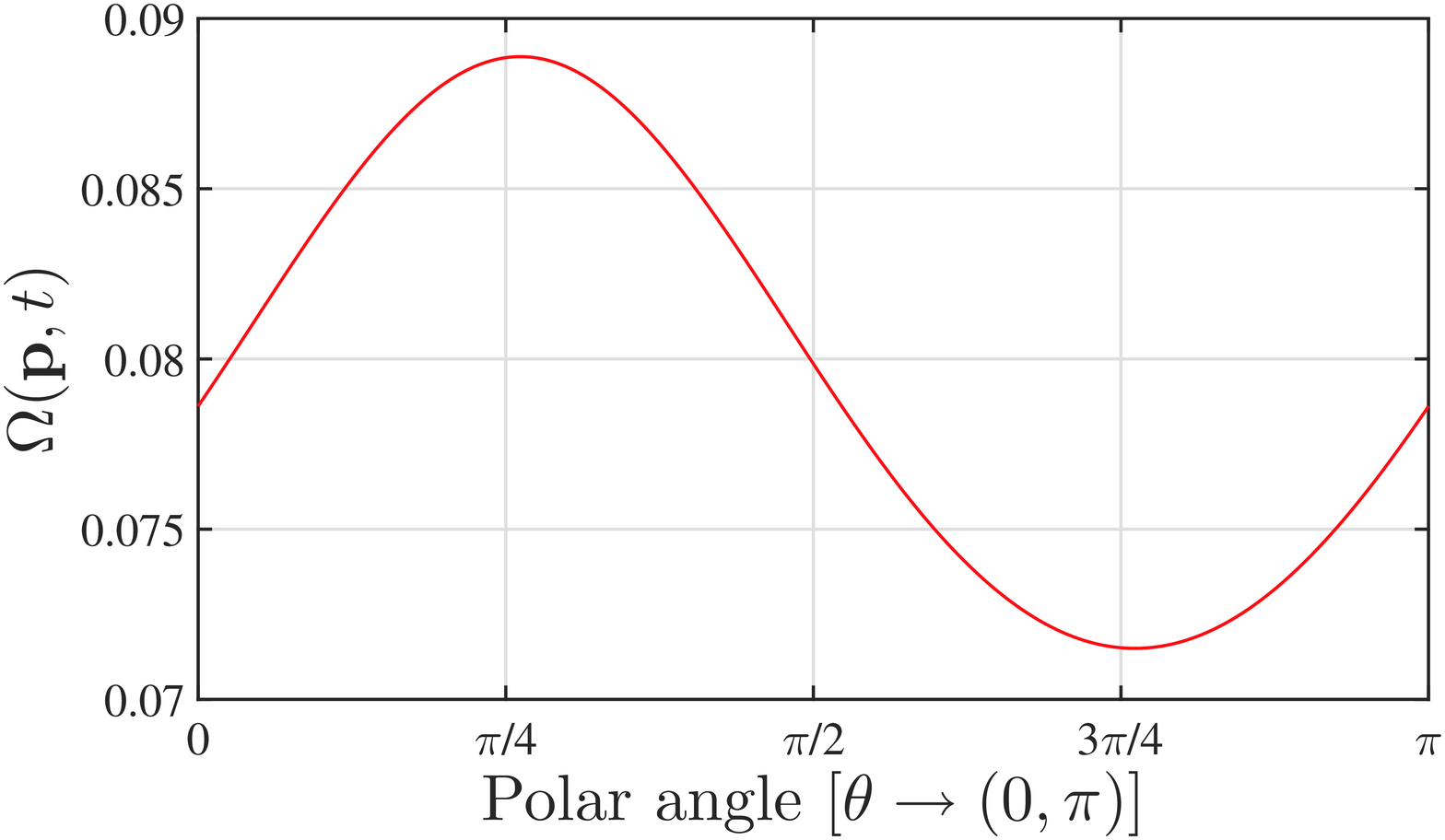}}
\subfigure[\hspace{-0.8cm}]{\includegraphics[scale=0.3]{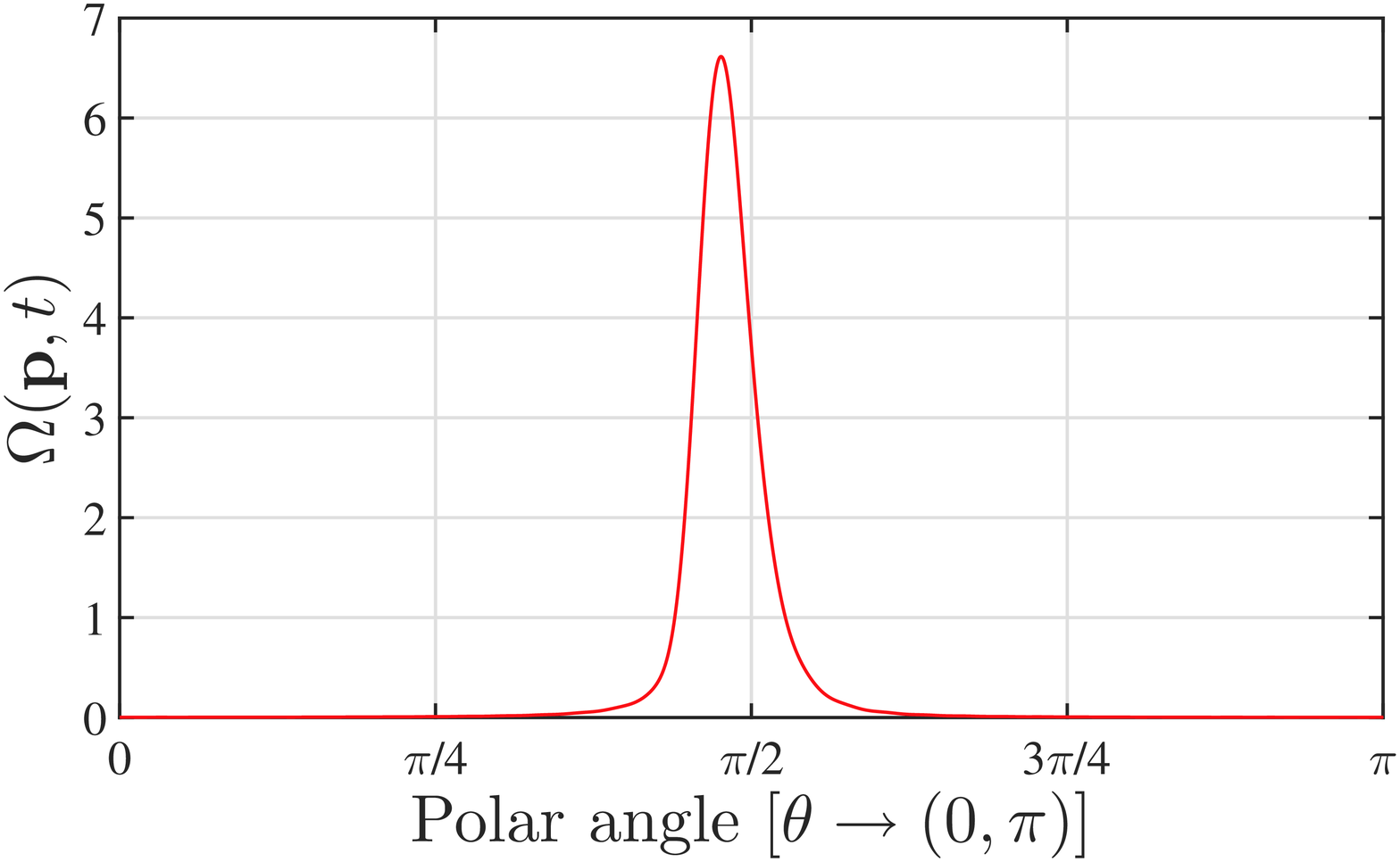}}
\end{center}
 \caption{Comparison of steady-state orientation probability density function ($\lim_{t\rightarrow\infty} \Omega(\boldsymbol{p}, t)$) along the flow-gradient plane for simple shear flow (a) $\Pen = 0.1$, (b) $\Pen = 100$.}
 \label{steady_distribution_comparison}
\end{figure}

In figure \ref{steady_distribution_comparison}, the steady-state probability density function ($\Omega(\boldsymbol{p})$) of the bacterium in the flow-gradient plane, for simple shear flow, for $\Pen =$ 0.1 and 100. It can be seen that at low $\Pen$, the distribution of swimmer orientations is such that it aligns close to the extensional axis, whereas at high $\Pen$, the peak of the probability density function is rotated towards the flow axis. 

\begin{figure}
\begin{center}
\subfigure[\hspace{-0.8cm}]{\includegraphics[scale=0.3]{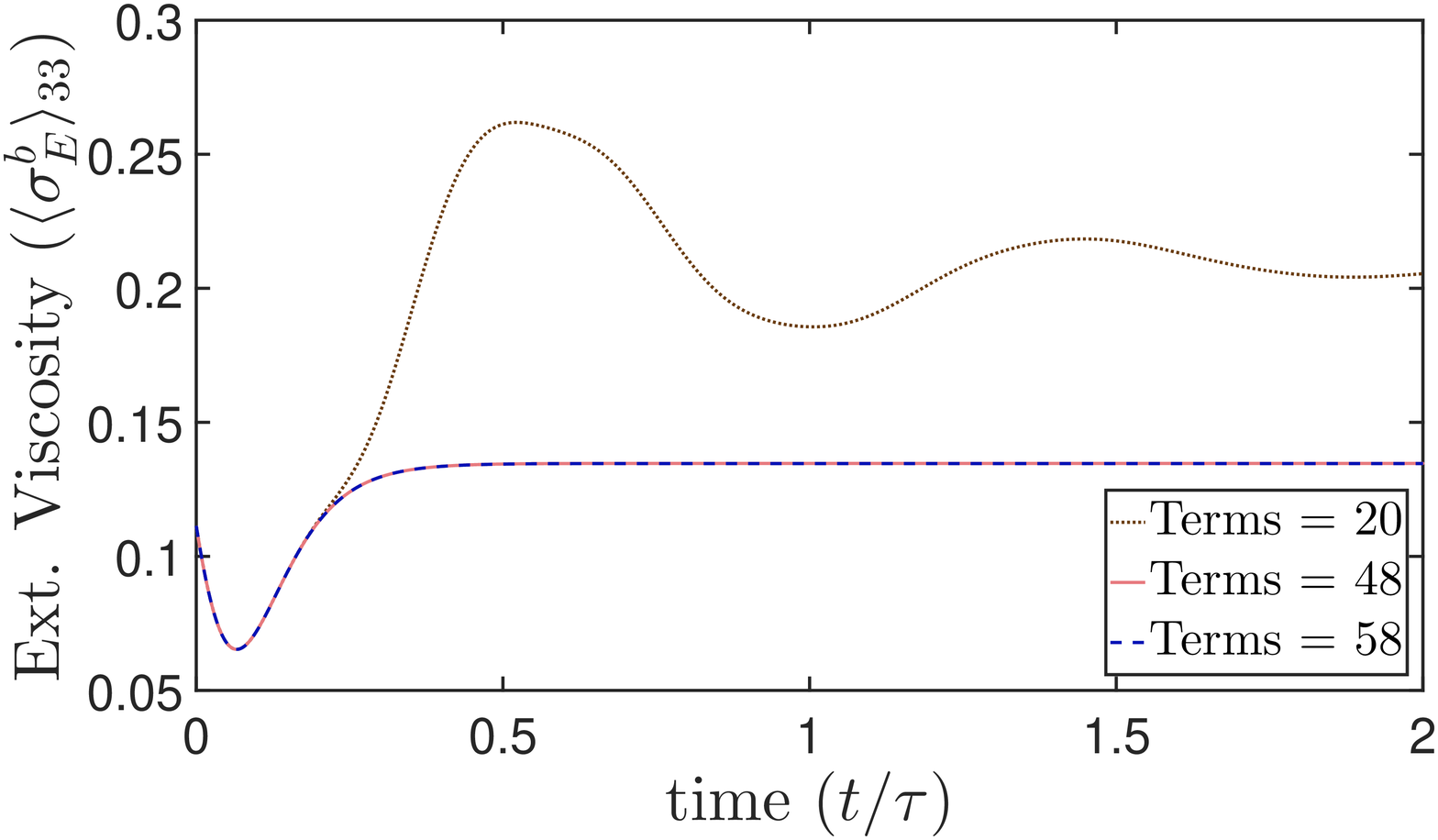}}
\subfigure[\hspace{-0.8cm}]{\includegraphics[scale=0.3]{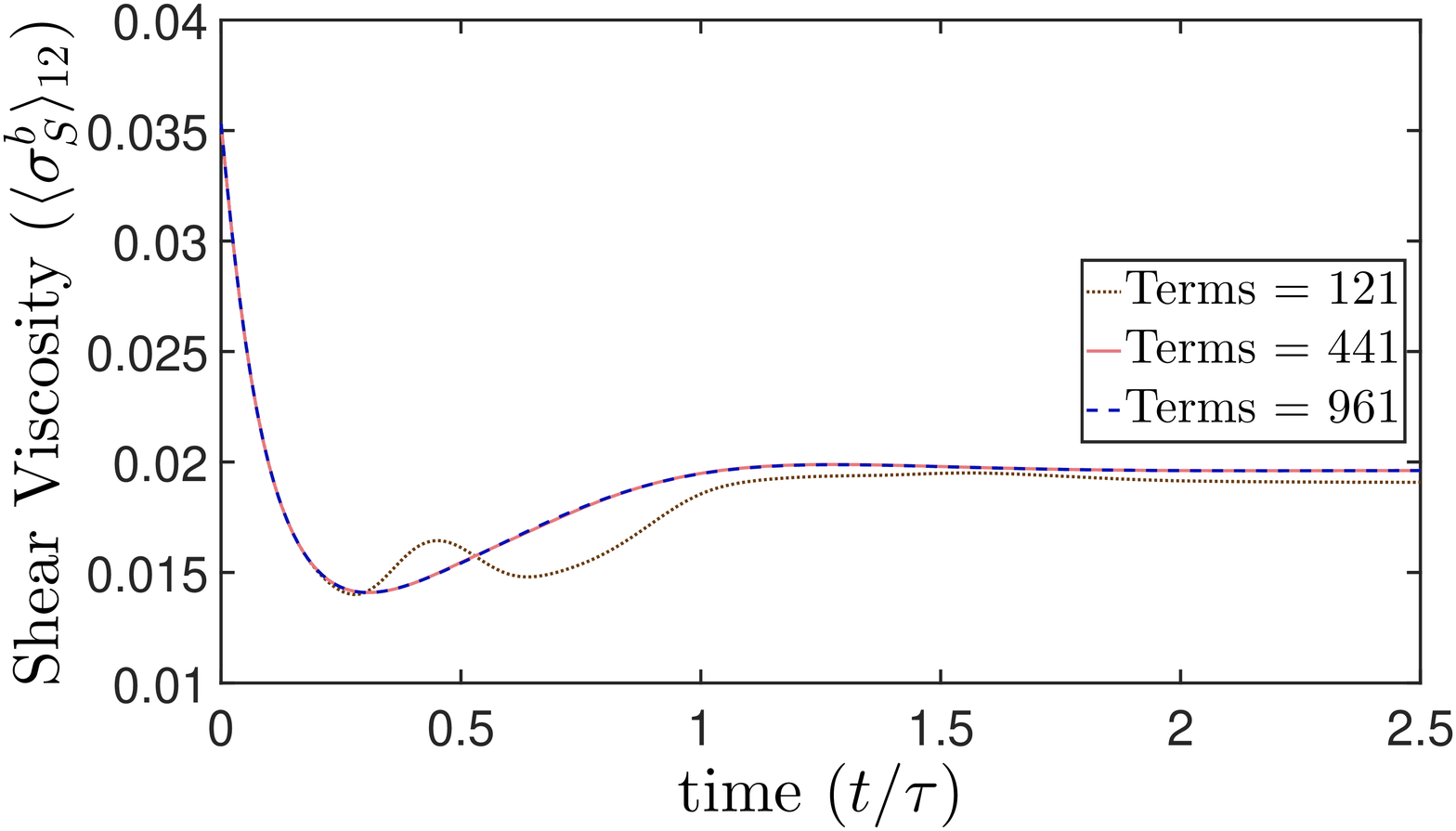}}
\end{center}
 \caption{Comparison of time dependent viscosity for different set of terms in the series expansion. The dotted curve (brown), the straight curve (light maroon) and the dashed curve (blue), corresponds to considering (a) 20, 48 and 58 terms for axisymmetric extensional flow and (b)121, 441 and 961 terms for simple shear flow, respectively in the series expansion of $\Omega(\boldsymbol{p}, t)$. }
 \label{stress_comparision_appendix}
\end{figure}

For both steady-state and transient responses, we conducted convergence tests with respect to the number of terms in the series expansion of $\Omega(\boldsymbol{p}, t)$. To validate the convergence at a given $\Pen$, we repeated the numerical calculations for various $l_{max}$ until the steady-state values of the rheological properties fell below a prescribed tolerence of 10$^{-6}$. We used the steady-state results of \cite{Saintillan2010, saintillan2010extensional} as the benchmark. In figure \ref{stress_comparision_appendix}, we have plotted the time dependent viscosity for $\Pen = 10$ for three values of $l_{max}$, viz. 20, 48 and 58 for axisymmetric extensional flow and 10, 20 and 30 for simple shear flow. In the latter case, it corresponds to considering 121, 441 and 961 terms in the series, respectively. Here, $D_r \tau = 0.062$ and the tumbles have been assumed to be random ($\beta=0$). It can be seen that the viscosity curves corresponding to the latter two values are coincident for both imposed 
flows. With fewer number of 
terms in the expansion, the transient response is oscillatory and the steady-state plateau shows significant offset. Similar behavior was reported by \cite{strand1987} while evaluating the transient rheological properties of the suspensions of rigid rods in simple shear.

\end{appendix}

\bibliographystyle{jfm}

\bibliography{references}

\end{document}